\documentclass[12pt,a4paper]{article}

\usepackage{journalstyle}

\usepackage[style=nejm, citestyle=numeric-comp, sorting=none]{biblatex}
\title{Joint-Embedding Predictive Architecture for Sensor-based Activity Recognition}

\author[1,3*]{Mohd Halim Mohd Noor}
\author[2]{Abdulrahman M. A. Baraka}

\affil[1]{School of Computer Sciences, Universiti Sains Malaysia, 11800 Pulau Pinang, Malaysia}
\affil[2]{Al-Quds Open University, Ramallah, Palestine}
\affil[3]{Artificial Intelligence Research Center, Ajman University, Ajman, United Arab Emirates}
\affil[*]{Address correspondence to: halimnoor@usm.my}

\date{}

\begin{document}

\maketitle

\begin{abstract}
Sensor-based human activity recognition (HAR) has achieved significant progressed in fully supervised learning settings. However, these supervised learning models rely on large amount of labeled data, which require labor-intensive collection and meticulous annotation. To address these challenges, this paper proposes a Joint Embedding Predictive Architecture framework tailored for sensor-based HAR, designed to learn robust and generalizable representations from unlabeled datasets. The proposed framework features an encoder designed to explicitly model both the fine-grained local temporal representations within individual window and the long-term temporal sequence of adjacent windows. Furthermore, we introduce an improved Variance-Invariance-Covariance Regularization (VICReg) objective function that incorporates computationally lightweight norm term to stabilize the JEPA pre-training phase. This term balances variance, invariance and covariance constraints to prevent representation collapse. The proposed HAR-JEPA framework is evaluated using two benchmark continuously performed activity datasets. The results show that high-quality representations are successfully learned by the proposed framework. Furthermore, the representations learned by HAR-JEPA demonstrates superior generalization on minority, high variance transitional activities such as sit-to-stand and sit-to-lie where supervised learning tend to overfit due to limited support.

\vspace{1em}
\noindent \textbf{Keywords:} Self-supervised learning, representation learning, human activity recognition, sensor-based HAR
\end{abstract}


\section{Introduction}
Human activity recognition using wearable sensors is an essential component in modern health monitoring \cite{Maddala2026}, human behavior analysis \cite{Etumusei2025} and sports science \cite{Zhang2026}. Traditionally, the field has relied heavily on supervised learning, whereby inertial sensors are used to collect motion data from human subjects performing a series of activities, and the data are meticulously labeled on row-by-row basis. Then, the machine learning and deep learning algorithms are used to model the activity data for activity classification. However, the process of labeling the data is time consuming, expensive and prone to errors, and this remains a significant obstacle to scaling these models across diverse populations and activities \cite{Haresamudram2025}.

To mitigate this, self-supervised learning (SSL) has emerged as a powerful alternative. In SSL, predictive models are built without relying on labeled datasets, instead they leverage large amounts of unlabeled data by exploiting the intrinsic structure of the data through pretext tasks to learn meaningful representations that can be transferred to downstream tasks. In this approach, the model training is divided into two stages. First, the models are trained to solve a pretext task that involves unlabeled data. Then the same models are fine-tuned on downstream tasks using labeled datasets. During pre-training, the activity signals are first transformed into different augmented versions, and the models are trained to recognize these transformations. The signals maybe masked or partially corrupted, and the models are trained to reconstruct the missing segments, encouraging them to capture the underlying temporal dynamics and contextual dependencies within the sensor data \cite{yuan_self_supervised_2024, logacjov_selfpab_2024}. 

Another SSL approach is contrastive learning, in which models learn useful features by comparing unlabeled data and identifying similarities and differences. For sensor-based HAR, multiple augmented views of the same segmented activity signals are generated through time shifting, scaling, jittering and masking. Views derived from the same segment are treated as positive pairs, while views derived from different segments are treated as negative pairs. The models are then trained to minimize the distance between positive pairs and maximize the distances between negative pairs in the latent space \cite{chen_temporal_2025}. To further incorporate domain-specific inductive biases within the model, \cite{huang_tfc_2025} introduces a learnable Fourier layer that produces time-frequency representations, and formulates instance-level contrastive loss in frequency domain. 

Although SSL approaches are effective, both pretext-based and contrastive SSL approaches have inherent limitations. Pretext methods rely on manually designed surrogate tasks which many not fully capture the semantic or temporal structures of sensor data, leading to suboptimal representations. In contrast, contrastive methods require large batches of negative samples and carefully constructed pairs to prevent representation collapse. Furthermore, contrastive learning operates at the actual instance similarity, encouraging models to distinguish between individual samples rather than explicitly model the underlying semantic or temporal structure of the data, which limits its ability to model more abstract and predictive representations.

Recently, a novel approach to SSL was introduced that shifts the learning objective from instance discrimination to predictive representation learning. The novel approach, termed Joint-Embedding Predictive Architecture (JEPA) learns by predicting latent representations of masked or future observations from contextual information in the latent space. A JEPA model consists of a context encoder, a target encoder and a predictor. The context encoder produces feature representation from partially visible data, while the target encoder generates target representation from the complete input. The predictor then uses the context features to predict these target representations. This predictive task enables learning directly in latent representation space and encourages the model to capture high-level semantic and temporal dependencies while avoiding negative samples. Furthermore, by operating in the latent space, the model can ignore unpredictable, high-entropy and low level details and instead focus on structurally meaningful information that better reflects the underlying dynamics of the data. As a result, JEPA produces more abstract, robust, and transferable representations that are particularly well-suited for complex sensor and time-series modeling tasks. 

In computer vision, JEPA has been employed to predict the hidden content from visible context in the latent space. Notably, the image-based JEPA (I-JEPA) leverages context–target masking strategies to predict the embeddings of masked image regions from surrounding visible regions, enabling the model to learn semantically meaningful and spatially coherent representations without reconstructing raw pixels \cite{assran_self_supervised_2023}. Video-JEPA (V-JEPA) extends JEPA to the video domain by learning to predict latent representations of masked spatiotemporal blocks from visible context in video clips \cite{bardes_revisiting_2024}. V-JEPA models are trained solely with a feature prediction objective on large-scale unlabeled video data, leading to versatile representations that capture both motion and appearance information. Evaluations show that these learned representations perform competitively on downstream tasks such as image classification, action recognition and temporal understanding.

Inspired by these advances, we investigate JEPA-based self-supervised learning on human sensor data. We propose HAR-oriented Joint Embedding Predictive Architecture (HAR-JEPA) tailored for wearable inertial measurement unit data effectively addressing the aforementioned HAR challenges. To the best of our knowledge, this work is the first attempt at investigating the use of JEPA for sensor-based HAR. We first pre-train the encoder using JEPA, and subsequently fine-tune the learned models for activity classification under limited labeled data. For pre-training, we use the large-scale REALDISP Activity Recognition dataset, which contains multi-sensor recordings collected from wearable devices attached to multiple body locations, covering 33 activities across three different scenarios. We design two encoder variants: a convolutional neural network (CNN)-based architecture and a transformer-based architecture. 

Furthermore, to enhance the ability of the context encoders to extract salient temporal features, we introduce a window-level embedding mechanism that explicitly models the temporal sequence of adjacent windows. Specifically, each activity window is first divided into fixed-length sub-windows (patches) to obtain local temporal representations, after which learnable window embeddings are added to differentiate neighboring windows within a sequence. This design allows the encoder to capture both intra-window dynamics through feature extraction layers and inter-window temporal dependencies through window-aware embeddings. 

\section{Related Work}
In HAR, supervised deep learning models such as CNNs\cite{baraka_deep_2025}, LSTMs\cite{MohdNoor2022}, Bi-LSTM\cite{Ige2025}, and Autoencoder\cite{MohdNoor2021} have achieved outstanding success in recognizing basic activities (e.g., walking or sitting). 
Nevertheless, the performance of these models remains highly dependent on labeled datasets. Annotating the sensor data signals for human activities is particularly challenging, as labeling large volumes of data is costly, complex, and time-intensive. To mitigate this, Generative Adversarial Networks (GANs) have been adopted as a semi-supervised learning strategy, leveraging unlabeled or partially labeled data to improve performance by learning representations that generalize to unseen distributions\cite{Shi2021}. While GANs have shown promise in generating balanced and realistic synthetic sensor data, training robust GAN models remains difficult due to the wide variability in signal amplitude, frequency, and period across different activities\cite{Zhang2022}.

Self-supervised learning (SSL) techniques have also been explored to reduce reliance on labeled data in sensor-based HAR\cite{chen_temporal_2025}. However, real-world data collection introduces multiple sources of noise, such as electromagnetic interference or scheduling uncertainties in sampling devices, that degrade data quality\cite{Zhang2022}. Since SSL models rely directly on raw sensor inputs during training, such noise negatively impacts recognition accuracy, particularly for transitional activities. JEPA offers a potential solution by extracting essential motion patterns from raw sensor data while suppressing irrelevant noise. Unlike conventional approaches, JEPA explicitly learns to disregard details it identifies as noise, thereby enhancing robustness in activity recognition. Unlike generative models, the emergence of JEPA has shifted the focus of SSL from instance-level reconstruction and manual data augmentation towards predicting latent representations of masked data\cite{zhang2026}. 

Originally, JEPA was popularized in computer vision (I-JEPA and V-JEPA). Recently, researchers have begun exploring its approach in time-series and sensor-based domains, offering a promising path for Sensor-based HAR. One of the earliest work of JEPA for time series data is reported in \cite{fei_jepa_2024}. Their model A-JEPA demonstrates that the masked-modeling principle can effectively learn audio representations by making predictions in a latent space. However, applying JEPA to time series is generally more straightforward than to images or video because the structure of the data is simpler and sequential. Time series naturally follow a one-dimensional temporal order, so defining context and prediction targets is intuitive. To make the prediction more challenging, given the high correlation among adjacent samples, the authors employ a curriculum-style region masking of audio spectrograms, gradually shifting from random blocks to time-frequency-aware masking. The vision transformer (ViT) is used as the context and target encoders. A similar study is reported in \cite{tuncay_audio-jepa_2025}, which utilizes a vision transformer to predict latent representations of masked spectrogram patches. The work achieved performance comparable to established models like wav2vec 2.0 while using only 20\% of the training data. However, since the audio signals are converted to mel-spectrogram images, the studies effectively treat the problem as a vision task rather than operating directly on raw temporal signals.

JEPA has been investigated for electrocardiogram (ECG) classification \cite{weimann_self-supervised_2024} which shares many signal characteristics with HAR sensors such as accelerometers and gyroscopes. In ECG-JEPA, both the encoders and the predictor are based on standard ViT, with the predictor having half the embedding size of the encoder. The ECG signal is split into non-overlapping 1D patches, and several contiguous blocks of patches are masked for predicting the missing representations in a latent space. The authors use an iterative strategy to sample non-overlapping contiguos masking blocks, enforcing a minimum block size to prevent fragmentation and ensure consistent input shapes across batches. The experimental results show that ECG-JEPA outperforms other SSL approaches, achieving state-of-the-art performance. A similar work is reported in \cite{kim_learning_2024} in which the authors employ random masking and multi-block masking. TS-JEPA is a JEPA framework for general time series. The time series data is first transformed into a sequence of non-overlapping patches using a one-dimensional convolutional neural network. The sequence of patches are randomly masked, and the visible patches are fed to the transformer-based context encoder. The proposed TS-JEPA is evaluated on ECG, weather, electricity and fault detection datasets. The experimental results demonstrate that TS-JEPA achieves a strong performance in both classification and forecasting tasks. 

Although ECG-JEPA and TS-JEPA have shown strong performance in their respective domains, a study on JEPA for sensor-based HAR is still needed. First, the context encoder accepts a single time series window as input, converts it into a sequence of patches and produces the latent embeddings from visible patches for the predictor to predict the missing embeddings. While this allows the model to learn the underlying structure of the input data, it often overlooks the broader sequential context of human activities which are inherently sequential. For example, a standing activity is often followed by either walking or a stand-to-sit transition. Furthermore, sensor HAR data typically involves a much larger number of activity classes, ranging from simple activities and gestures to complex daily routines. 

Furthermore, JEPA framework operates in a high-level latent space by employing the regression prediction loss, formulated as Mean Squared Error or $L_2$ distance. While the loss function has demonstrated a strong performance in computer vision, it is often less effective for sensor HAR. In computer vision, neighboring image regions are highly spatial and semantic correlated, allowing the model to accurately predict latent representations of masked regions. Although activity signals are driven by large-scale body movements that are often repetitive (gait) or sparse (body posture), they are low-dimensional and inherently noisy, temporally ambiguous and highly variable across subjects. A context window may correspond to multiple possible future activities or motions, causing the $L_2$ objective to regress toward the average of multiple latent representations rather than learning discriminative features, ultimately leading to representation collapse. 

VICReg is a self-supervised objective framework that prevents representation collapse by introducing variance and covariance regularization alongside the invariance (distance) term  \cite{bardes_vicreg_2022}. Although the variance and covariance terms able to force the learned embeddings to spread out and remove redundancy, they are often struggle with scale instability in sensor data. To address this limitation, a flexible combination of Layer Normalization and weight decay is used to ensure stable pre-training convergence \cite{huang_bijepa_2026}. Radial-VICReg introduces a radial Gaussianization distribution that regulates the norm of embeddings to follow Chi distribution, leading to a more diverse representation \cite{kuang_radial-vcreg_2026}. Kernel VICReg allows the self-supervised model to operate in a non-linear latent space instead of Euclidean space by applying kernel transformation and optimizing the objectives using Hilbert-Schmidt norm \cite{sepanj_kernel_2025}. However, these approaches introduce either distribution-specific assumptions or additional computational complexity through kernel computation.

Therefore, this paper aims to investigate the use of JEPA approach for sensor-based HAR that models the temporal dependencies between two adjacent windows. Specifically, we design a context encoder architecture that not only models the sequence of patches but also the temporal sequence of the input windows. Furthermore, we propose two variants of context encoder that are based on convolutional neural network and transformer, respectively, to capture both local and global temporal patterns effectively. Furthermore, we introduce a norm term to VICReg objective that is computationally lightweight, directly regularize the magnitude of the embedding vector to stabilize the geometry of the embedding space. The proposed method has been evaluated on three HAR benchmark datasets: REALDISP for pre-training, SBHARPT and FORTH-TRACE for downstream human activity recognition. The results show that the proposed method outperform pure supervised learning-based activity recognition.

\section{Proposed Method}
\subsection{Overall Architecture}
\begin{figure}[t]
    \centering
    \includegraphics[scale=0.65]{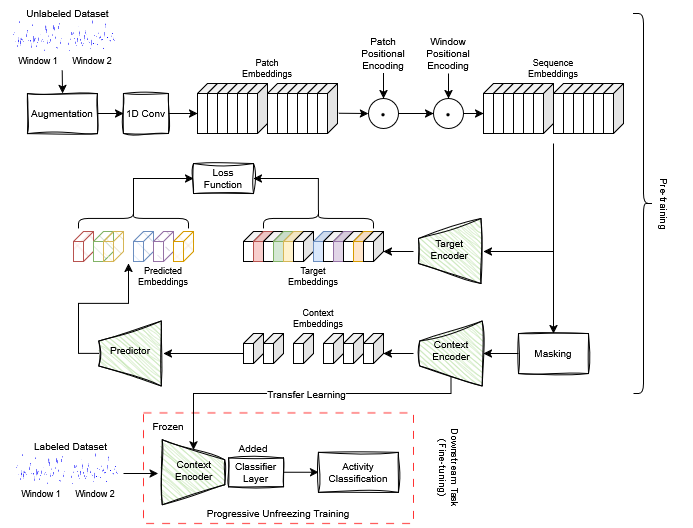}
    \caption{Overview of HAR-JEPA framework. The process is divided into two phases: a self-supervised pre-training phase on unlabeled sensor data, and a supervised fine-tuning phase for the downstream activity recognition task.}
    \label{fig:fig1_overall_har_jepa}
\end{figure}

This work presents a Joint-Embedding Predictive Architecture (JEPA) tailored for sensor-based HAR that learns transferable motion representations through self-supervised temporal prediction rather than direct signal reconstruction. Figure \ref{fig:fig1_overall_har_jepa} illustrates the overall HAR-JEPA framework which consists of two phases: pre-training and downstream activity classification. As shown in the figure, given two consecutive overlapping or non-overlapping windows are first sampled from the unlabeled dataset and enriched through an augmentation process. Activity signals are inherently noisy, highly correlated with the preceeding and subsequent samples, and dependent on the sensor's body placement. Simple masking is therefore insufficient to capture these long-range temporal dependencies and structured correlations across time, as it primarily enforces local semantic learning while ignoring the high-level semantic transitions and continuity across windows. 

We employ two complementary augmentation strategies to encourage the model to learn invariant and robust representations from activity signals. First, we apply Gaussian noise to the input signals, which mimics sensor inaccuracies and environmental interference, forcing the model to focus on the underlying motion patterns. Second, we randomly reverse the sensor channels to account for potential variations in device orientation and sensor placement. This strategy prevents the model from over-relying on specific coordinate axes, encouraging the model to learn more generalized and orientation-invariant features that are robust across different users and device positions.

The sensor data is then converted into a sequence of non-overlapping temporal patches. To this end, we adopt a 1D convolutional layer with kernel size and stride equal to patch size to convert raw sensor signal into a sequence of patches. Specifically, each window is divided into $N=\frac{L}{P}$ non-overlapping patches of length $P$, where $P$ is the patch size. Formally, given two concatenated consecutive windows $\mathbf{w} \in \mathbb{R}^{2 \times L \times C}$, where $L$ denotes the window size and $C$ the number of sensor channels, the patch embedding layer applies

\begin{equation}
\mathbf{Z} = \texttt{Conv1D}(\mathbf{w}) \in \mathbb{R}^{2 \times N \times D}
\end{equation}

where $D$ is the embedding dimension. Please note that the batch dimension is omitted for clarity.

To encode the temporal order of patches within a window, positional embeddings $\mathbf{E}_{\texttt{pos}} \in \mathbb{R}^{N \times D}$ are added to the patches. In the convolutional encoder, these embeddings are learnable parameters while fixed 1D sine-cosine positional encodings are used in the transformer-based encoder.

\begin{equation}
\mathbf{Z}_p = \mathbf{E} + \mathbf{E}_{\texttt{pos}}
\end{equation}

To explicitly model the temporal dependency between two adjacent windows, we introduce learnable window embeddings $\mathbf{E}_{\texttt{win}} \in \mathbb{R}^{2 \times D}$, where each index corresponds to a window position (previous or current). The corresponding embedding vector is broadcasted across the patch dimension and added to all patches of each window to produce window-aware patch embeddings as follows:

\begin{equation}
\mathbf{Z}_w = \mathbf{Z}_p + \texttt{Expand}(\mathbf{E}_{\texttt{win}})
\end{equation}

The embeddings allows the encoder to preserve absolute temporal information of the patches within the window and differentiate adjacent windows, encouraging the learning of temporal representations across windows. Upon adding the patch and window positions, a fixed number of patches are randomly masked. 

As shown in Figure \ref{fig:fig1_overall_har_jepa} the framework consists of three main networks: context encoder, target encoder and predictor. The context encoder takes the visible patches as input and generates context embeddings. The predictor takes the context embeddings and estimates the latent embeddings of masked patches while the target encoder encodes the original, unmasked patches to generate stable target representations for the masked regions, which the predictor is trained to match during learning. After self-supervised pre-training, the learned context encoder is transferred to downstream HAR tasks by attaching a simple classifier head and fine-tuning selected layers. This design enables efficient learning from large unlabeled datasets while maintaining strong performance in supervised activity recognition.

\subsection{Model Architecture}
To model temporal dependencies between two adjacent windows in the JEPA framework, we design an encoder architecture that operates on sequences of time series patches, that models both intra-window and inter-window temporal relationship. Two encoder variants are proposed: a convolutional encoder for efficient local temporal modeling and a transformer-based encoder which can effectively capture long-range dependencies.

\textbf{Convolutional Encoder}. The convolutional encoder is designed to efficiently capture local temporal dependencies within and across patches. It consists of a stack of convolutional blocks operating on the patch sequence. Each convolutional block contains two components: a depthwise temporal convolution and a channel-wise feedforward network. Since the depthwise convolution is applied independently to each embedding channel, it models temporal relationships across neighboring patches without mixing channel information. The channel-wise feedforward network expands the embedding dimension by a factor of four, applies a GELU activation, and projects it back to the original dimension. This structure is similar to the multilayer perceptron block in Transformers and enables non-linear channel mixing. The output is then added to the input through a residual connection. The $l$-th convolutional block can be defined as follows:

\begin{equation}
\mathbf{Z}^{l+1}_w = \mathbf{Z}^l_w + f_c(f_d(\mathbf{Z}^l_w))
\end{equation}

where $f_d$ denotes the depthwise convolution and $f_c$ denotes the channel-wise feedforward network.

The stack has varying kernel sizes, allowing the model to capture multi-scale temporal patterns. Larger kernels focus on broader temporal contexts, while smaller kernels capture finer local features. The final feature maps are then passed through layer normalization, producing the latent representations used in the JEPA objective.

\textbf{Transformer-based Encoder}. We propose a transformer-based encoder that captures the global temporal dependencies between patches. To retain patch ordering information, we incorporate fixed sinusoidal positional embeddings. Unlike the traditional transformer encoder that processes each window independently, the proposed encoder incorporates window embeddings to explicitly model the sequence of adjacent windows as described above. After adding both positional and window embeddings, the patches are passed through a Transformer stack, where each layer comprises multi-head self-attention and a feed forward network, with residual connections and layer normalization applied to each sublayer. This design enables the model to not only capture long-range temporal dependencies within the window but also across adjacent windows, which particularly essential for activities with transitional dynamics. A final layer normalization is applied to produce the output representations. Table \ref{tab:summary_encoder} provides the summary of the proposed encoders.

\begin{table}[t]
    \caption{Summary of the main components of the proposed encoders.}    
    \centering
    \small
    \begin{tabularx}{\linewidth}{lXX}
            \hline
            Component & Convolutional Encoder & Transformer-based Encoder \\  
            \hline
            Patch embedding & 1D Conv (kernel=stride=$P$) & 1D Conv (kernel=stride=$P$)\\ 
            Positional encoding & Learnable embedding & Fixed 1D sine-cosine \\
            Window encoding & Learnable embedding & Learnable embedding \\
            Core blocks & Depthwise conv + FFN$^1$ & Multi-head self-attention + FFN$^1$ \\
            Normalization & Layer norm & Layer norm \\
            Dependency range & Local to mid-range (kernel-controlled) & Global (self-attention) \\
            \hline
            \multicolumn{3}{l}{\footnotesize $^1$ FFN: Feedforward Network} \\
    \end{tabularx}

    \label{tab:summary_encoder}
\end{table}

\textbf{Predictor}. We design a lightweight transformer-based predictor that estimates the latent representations of masked patches from the context embeddings generated by the context encoder. We deliberately limit the predictor's Transformer stack to half the depth of the encoder, encouraging the encoder to capture more robust, high-level features during the self-supervised learning phase. Given the context embeddings, the predictor first maps them from the encoder embedding dimension to a lower-dimensional predictor space through a linear projection layer. This dimensionality reduction improves training stability and reduces computational complexity while preserving the semantic structure required for prediction. Then, fixed sinusoidal positional embeddings are added to retain patch ordering information. 

For prediction of masked patches, a learnable mask token is replicated to match the number of target patches, and augmented with their corresponding positional embeddings and concatenated with the context embeddings along the sequence dimension. This concatenated sequence is processed by a Transformer stack which utilizes multi-head self-attention to model the dependencies between the context and target embeddings. The final feature maps are then passed through layer normalization, and mapped back to the original embedding dimension, ensuring that the predicted representations reside in the same latent space as the targets.

\subsection{Pre-training}
The pre-training objective of HAR-JEPA is to minimize the distance between the predicted representation and the target representation. Specifically, we employ a multi-term objective function based on Variance-Invariance-Covariance Regularization (VICReg) framework \cite{bardes_vicreg_2022}. The loss function is defined as a weighted sum of four terms: invariance $L_s$, variance $L_v$, covariance $L_c$ and norm $L_n$ terms. Let $Z_{p}$ denotes the predicted representation produced by the predictor network and $Z_{t}$ the target representation produced by the target encoder, the total loss is defined as


\begin{equation}
L(Z_{p}, Z_{t}) = \nu_{s} L_{s}(Z_{p}, Z_{t}) + \nu_{v} (L_{v}(Z_{p}) + L_{v}(Z_{t})) + \nu_{c} (L_{c}(Z_{p}) + L_{c}(Z_{t})) + \nu_{n} L_{n}(Z_{p})
\end{equation}

where $\nu_{s}, \nu_{v} $, $\nu_{c}$ and $\nu_{n}$ are hyper-parameters for scaling the invariance, variance, covariance and norm terms respectively. 

The invariance term $L_{s}$ encourage the predicted representation to match the target representation. Instead of mean square error, a Smooth L1 (Huber) loss is used as follows:

\begin{equation}
L_{s}(Z_p, Z_t) = \frac{1}{N} \sum_{i=1}^{N} \texttt{Smooth L1}(Z_{p}, Z_{t})
\end{equation}

The variance term prevents the encoder from collapsing to a single point by enforcing a minimum standard deviation across the batch for each embedding dimension. Given a representation matrix $Z \in \mathbb{R}^{M \times D}$ where $M$ is the product of batch size and sequence length, and $D$ is the embedding dimension, we compute the standard deviation of each embedding dimension across the combined batch and temporal dimensions:

\begin{equation}
\sigma_j = \sqrt{\frac{1}{M-1} \sum_{i=1}^{M} (Z_{i,j} - \bar{Z}_j)^2 + \epsilon}
\end{equation}

where $\bar{Z}_j$ is the mean of $j$-th dimension. The variance loss is defined as follows:

\begin{equation}
L_{v}(Z) = \frac{1}{N} \sum_{i=1}^{N} \textbf{max}(0, \gamma - \sigma_i)
\end{equation}

where $\gamma$ is the predefined threshold that enforces a minimum standard deviation of the predicted and target representations.

The covariance term reduces redundancy in the learned embeddings by encouraging decorrelation. Given a representation matrix $Z \in \mathbb{R}^{M \times D}$ where $M$ is the product of the batch size and sequence length, and $D$ is the embedding dimension, we first center and normalize the representations:

\begin{equation}
\hat{Z}_{i,j} = \frac{Z_{i,j} - \mu_j}{\sigma_j + \epsilon}
\end{equation}

where $\mu_j$ and $\sigma_j$ are the mean and standard deviation of dimension $j$ across the batch. The covariance matrix is then computed as follows:

\begin{equation}
C = \frac{\hat{Z}^{\top}\hat{Z}}{N-1}
\end{equation}

The covariance regularization is defined as the sum of squared off-diagonal elements of $C$:

\begin{equation}
L_c(Z) = \frac{1}{D} \sum_{i \neq j} C_{i,j}^2
\end{equation}

Although the variance and covariance terms able to force the learned embeddings to spread out and remove redundancy, they are often struggle with scale instability in sensor data. To ensure the predicted embeddings remain within a stable numerical range, which is important for sensor-based HAR where the embedding magnitude may vary significantly, we introduce a norm regularization term. The term prevents the embeddings from shrinking into a single point (collapse to origin). Specifically, the term penalizes the squared deviation of the embedding norm from a target value $\tau$.

\begin{equation}
L_{n}(Z_p) = \frac{1}{N} \sum_{i=1}^{N} (\Vert Z_i \Vert_2 - \tau)^2
\end{equation}

where $\tau = 1 + \sigma (\sqrt{d} - 1)$, with $d$ is the embedding dimension and $\sigma$ is the dispersion coefficient that determines how far the target norm is pushed beyond the unit sphere relative to the embedding dimension $d$. 

For pre-training, we use the AdamW optimizer \cite{loshchilov_decoupled_2019} with $\beta_1 = 0.9$, $\beta_2 = 0.99$ and $\epsilon = 1e-8$ over 200 epochs. The learning rate follows a linear warp-up for the first 10\% of training steps, increasing from a base learning rate of $1 \times 10^{-4}$ to a maximum learning rate of $1 \times 10^{-3}$, and then decays to zero using a cosine schedule for the remaining training steps. The target encoder is updated using exponential moving average of the context encoder's weights. The momentum coefficient is progressively increased during training, starting from $0.99$ to $0.996$ in the first 10\% of training steps, followed by a cosine schedule that increases it to $0.9995$ for the remainder of training. 

In the experiments, we set the hyperparameter of each term as follows: $\nu_{s} = 1.0$, $\nu_{v} = 5.0$, $\nu_{c} = 0.2$ and $\nu_{n} = 0.1$. For the variance term, $\gamma = 0.5$ is selected because sensor data exhibits less variability compared to high-dimensional unstructured data such as images and videos. Finally, for the proposed norm term, we evaluate three different values of $\sigma$ (0.20, 0.25 and 0.30) to study its influence on the dispersion of learned representations. The JEPA model is trained using REALDISP dataset which we fully describe in Section \ref{dataset_description}.

\subsection{Downstream Task: Human Activity Recognition}\label{downstream_har}
After JEPA pre-training, the context encoder is retained, while the target encoder is discarded. The context encoder is then added with a classification head. Specifically, the patch embeddings generated by the encoder are concatenated across views and aggregated via max pooling along the patch dimension. The classification head consists of a fully connected hidden layer with a GELU activation function followed by a final predictive layer. The the hidden layer size is set to half of the embedding size, and dropout regularization is applied after the hidden layer. The drop rate is set to 0.3. The encoder is then fine-tuned and evaluated on sensor-based activity classification tasks. 

Instead of implementing a conventional full network fine-tuning, we adopt a structured, progressive unfreezing training strategy to preserve the low-level structural features learned during JEPA pre-training. The unfreezing schedule is executed in four distinct phases throughout the training: 

\begin{itemize}
    \item Phase 1 (epochs 1-10): The context encoder is frozen, restricting gradient updates to the classification head only.
    \item Phase 2 (epochs 11-40): The final encoder (convolution or transformer) block, together with the layer normalization layer are unfrozen, allowing the highest semantic layer to adapt to the target activity classes.
    \item Phase 3 (epochs 41-70): The top two encoder blocks are unfrozen, allowing deeper adaptation of task-specific features.
    \item Phase 4 (epochs 71-100): The remaining encoder blocks are fully unfrozen, allowing the entire encoder to be fine-tuned end-to-end. This stage refines intermediate and low-level feature representations to maximize downstream classification performance.
\end{itemize}

The models are trained using AdamW optimizer with $\beta_1 = 0.9$, $\beta_2 = 0.99$ and $\epsilon = 1e-8$ over 100 epochs. Discriminative learning rates are used during the progressive training and fine-tuning, whereby newly unfrozen encoder blocks are assigned smaller learning rates than the classification head based on a decay factor of 0.975 as follows:

\begin{equation}
\eta_{l} = \eta_{b} \cdot 0.975^{l}
\end{equation}

where $\eta_{b}$ is the base learning rate which is set to 0.001 and $l$ denotes the encoder block level, where $l=1$ corresponds to the final encoder block and larger values correspond to lower encoder blocks. Furthermore, a cosine annealing schedule is applied throughout training to decrease all learning rates from their initial value to a minimum learning rate of $1 \times 10^{-6}$. Two benchmark datasets are used in the experiments which are described in Section \ref{dataset_description}.

\section{Experimental Setup}

\subsection{Dataset Description and Preprocessing} \label{dataset_description}
To evaluate the performance of the proposed architecture, three public benchmark datasets are utilized: REAListic sensor DISPlacement (REALDISP) \cite{realdisp} for pre-training, the Smartphone-Based Recognition of Human Activities and Postural Transitions (SBHARPT) \cite{sbharpt} and FORTH-TRACE \cite{4th} datasets for downstream tasks.

The REALDISP dataset was originally collected to examine how sensor displacement affects activity recognition in real-world scenarios.  The ideal and mutual-displacement settings represent extreme cases, serving as boundary conditions for recognition algorithms. In contrast, self-placement reflects how users might naturally attach sensors in everyday contexts, such as sports or lifestyle applications.
The dataset encompasses a diverse set of physical activities (including warm-ups, cool-downs and fitness exercises), multiple sensor modalities (acceleration, angular velocity, magnetic field, and quaternions), and recordings from 17 participants.

The SBHARPT dataset comprises recordings of 30 volunteers aged between 19 and 48 years. Each participant followed a protocol consisting of six basic activities: three static postures (Activity 3: Sitting, Activity 4: Standing, and Activity 5: Lying down) and three dynamic movements (Activity 0: Walking, Activity 1: Walking upstairs and Activity 2: Walking downstairs). In addition, six postural transitions between static postures were recorded: Activity 6: Stand-to-Sit, Activity 7: Sit-to-Stand, Activity 8: Sit-to-Lie, Activity 9: Lie-to-Sit, Activity 10: Stand-to-Lie, and Activity 11: Lie-to-Stand. During the sessions, participants wore a smartphone secured at the waist. The device’s embedded accelerometer and gyroscope captured triaxial acceleration and angular velocity at a constant sampling rate of 50 Hz. 

The FORTH-TRACE dataset was developed for human activity recognition, focusing on both basic activities and postural transitions. Data were collected from 15 participants, each equipped with five wearable sensor nodes placed at specific body locations. The participants performed 16 activities, comprising seven basic activities and nine postural transitions. The dataset includes “talking” activities, which overlap with movement such as talking while climbing stairs versus climbing stairs alone. Since accelerometer and gyroscope sensors are designed to capture physical motion rather than speech, this study treats such overlapping cases as a single activity (climbing stairs), consistent with prior works \cite{baraka_similarity_2023, baraka_deep_2025}. Accordingly, the experiments focus on four basic activities: Activity 0: Standing, Activity 1: Sitting, Activity 2: Walking, and Activity 3: Walking upstairs/downstairs, and two transitional activities: Activity 4: Stand-to-Sit and Activity 5: Sit-to-Stand. Both downstream task datasets are split into training, validation, and testing subsets as listed in Table \ref{tab:dataset}.

\begin{table}[t]
    \caption{Spliting data for both SBHARPT and FORTH-TRACE datasets}    
    \centering
    \small
    \begin{tabular}{cccc}
            \hline
            Dataset & Training & Validation & Testing \\  
            \hline
            SBHARPT & 70\% & 5\% & 25\% \\ 
            FORTH-TRACE & 70\% & 5\% & 25\%\\
            \hline
            \end{tabular}
    \label{tab:dataset}
\end{table}

The datasets undergoes a series of preprocessing steps, including cleaning and standardization. Data cleaning eliminates errors, duplicates, and inconsistencies to preserve integrity, while standardization adjusts feature value scales to ensure consistency and support stable model training. Signal segmentation is then performed using a fixed-size sliding window with a degree of overlap, enabling continuous signals to be partitioned into temporal segments. A uniform window size of 102 samples (about two seconds) is applied across all datasets, as it is sufficient for recognizing basic and transitional activities \cite{Straczkiewicz2021}.

For pre-training, a degree of 10\% overlap was used to ensure minimal redundancy between adjacent segments while still capturing temporal patterns in the sensor signals. This relatively low overlap helps increase the diversity of training samples and reduce computational overhead during the pre-training phase. The proposed HAR-JEPA accepts two sequential sensor windows as input, enabling the model to learn both discriminative representations within each window and the temporal relationship between consecutive windows. 

For the downstream human activity recognition task, the same two-window input configuration and window size are retained to preserve the learned temporal representations. The overlap is increased to 50\%, which is commonly adopted in human activity recognition as it provides a practical balance between capturing temporal dynamics and maintaining computational efficiency. 

\subsection{Evaluation Metrics}
To evaluate the recognition performance of the HAR model, standard metrics commonly employed in related studies were adopted, including accuracy, precision, recall, and F1-score. These metrics collectively provide a comprehensive assessment of the model's performance. The metrics are defined as follows:

\begin{equation}
\texttt{Accuracy} = \frac{(\texttt{TP} + \texttt{TN})}{(\texttt{TP} + \texttt{TN} + \texttt{FP} + \texttt{FN})}
\end{equation}

\begin{equation}
\texttt{Precision} = \frac{\texttt{TP}}{(\texttt{TP}+\texttt{FP})}
\end{equation}

\begin{equation}
\texttt{Recall} = \frac{\texttt{TP}}{(\texttt{TP}+\texttt{FN})}
\end{equation}

\begin{equation}
\texttt{F1-score} = \frac{2 \times \texttt{Precision} \times \texttt{Recall}}{\texttt{Precision} + \texttt{Recall}}
\end{equation}

where $\texttt{TP}$ is true positives, $\texttt{FP}$ is false positives, $\texttt{TN}$ is true negatives, and $\texttt{FN}$ is false negatives.

\section{Results and Discussion}
This section examines the effectiveness of the proposed HAR-JEPA framework in terms of linear probing and downstream activity recognition. Furthermore, the learned latent embeddings of the JEPA models are visualized and analyzed using t-distributed Stochastic Neighbor Embedding. t-SNE is nonlinear dimensionality reduction technique widely used to project high-dimensional data into lower-dimensional spaces for intuitive visualization. Finally, we evaluate the effect of norm term's dispersion coefficient of the improved VICReg loss function.

\subsection{Linear Probing Evaluation}
The quality of the learned latent embeddings by the proposed framework is evaluated through linear probing experiments across the two encoders on SBHARPT and FORTH-TRACE datasets. In linear probing, the pre-trained encoders are frozen, and only a linear classifier is trained on top of the learned latent representations to assess their discriminative capability without updating the encoder parameters. The experiments demonstrate that both the convolutional and transformer-based JEPA architectures successfully capture meaningful representations. 

The convolutional encoder demonstrates superior performance, achieving average F1-scores of 0.7183 and 0.6442, noticeably outperforming transformer-based encoder which achieves 0.6717 and 0.5196 for both datasets as shown in Table \ref{tab:avg_f1score_linear_probing}. Tables \ref{tab:linear_probe_conv_clf_report} and \ref{tab:linear_probe_xformer_clf_report} show the classification report for both encoders across SBHARPT and FORTH-TRACE datasets.

\begin{table}[t]
    \caption{Average F1-score measure of both encoders across SBHARPT and FORTH-TRACE datasets}    
    \centering
    \small
    \begin{tabular}{ccc}
            \hline
            Dataset & Convolutional Encoder & Transformer-based Encoder \\  
            \hline
            SBHARPT & 0.7183 & 0.6442\\ 
            FORTH-TRACE & 0.6717 & 0.5196\\
            \hline
            \end{tabular}
    \label{tab:avg_f1score_linear_probing}
\end{table}

\begin{table}[t!]
    \caption{Linear probing classification performance of the convolutional encoder across SBHARPT and FORTH-TRACE datasets}    
    \centering
    \small
    \begin{tabular}{ccccc}
            \hline
            Activity & Precision & Recall & F1-score & Support\\  
            \hline
            &&SBHARPT \\
            \hline
            Walking & 0.67 & 0.86 & 0.75 & 578\\ 
            Walking Upstairs & 0.90 & 0.85 & 0.87 & 566\\ 
            Walking Downstairs & 0.83 & 0.65 & 0.73 & 536\\ 
            Sitting & 0.82 & 0.91 & 0.86 & 666\\ 
            Standing & 0.89 & 0.82 & 0.86 & 679\\ 
            Lying down & 0.98 & 0.99 & 0.99 & 706\\ 
            Stand-To-Sit & 0.77 & 0.50 & 0.61 & 48\\ 
            Sit-To-Stand & 0.78 & 0.69 & 0.74 & 36\\ 
            Sit-To-Lie & 0.72 & 0.48 & 0.58 & 58\\ 
            Lie-To-Sit & 0.48 & 0.63 & 0.54 & 46\\ 
            Stand-To-Lie & 0.57 & 0.65 & 0.61 & 51\\ 
            Lie-To-Stand & 0.74 & 0.36 & 0.49 & 55\\
            \hline
            &&FORTH-TRACE \\
            \hline
            Standing & 0.87 & 0.67 & 0.76 & 842\\
            Sitting & 0.74 & 0.98 & 0.84 & 827\\
            Walking & 0.85 & 0.88 & 0.86 & 1396\\
            Up/Downstairs & 0.80 & 0.75 & 0.78 & 676\\
            Stand-To-Sit & 0.68 & 0.29 & 0.40 & 59\\
            Sit-To-Stand & 0.89 & 0.24 & 0.38 & 66\\
            \hline
            \end{tabular}

    \label{tab:linear_probe_conv_clf_report}
\end{table}

\begin{table}[t!]
    \caption{Linear probing classification performance of the transformer-based encoder across SBHARPT and FORTH-TRACE datasets}    
    \centering
    \small
    \begin{tabular}{ccccc}
            \hline
            Activity & Precision & Recall & F1-score & Support\\  
            \hline
            &&SBHARPT \\
            \hline
            Walking & 0.62 & 0.92 & 0.74 & 578\\ 
            Walking Upstairs & 0.88 & 0.81 & 0.84 & 566\\ 
            Walking Downstairs & 0.83 & 0.51 & 0.63 & 536\\ 
            Sitting & 0.64 & 0.91 & 0.75 & 666\\ 
            Standing & 0.84 & 0.71 & 0.77 & 679\\ 
            Lying down & 0.97 & 0.77 & 0.86 & 706\\ 
            Stand-To-Sit & 0.65 & 0.35 & 0.46 & 48\\ 
            Sit-To-Stand & 0.75 & 0.67 & 0.71 & 36\\ 
            Sit-To-Lie & 0.63 & 0.29 & 0.40 & 58\\ 
            Lie-To-Sit & 0.52 & 0.74 & 0.61 & 46\\ 
            Stand-To-Lie & 0.46 & 0.61 & 0.52 & 51\\ 
            Lie-To-Stand & 0.69 & 0.33 & 0.44 & 55\\ 
            \hline
            &&FORTH-TRACE \\
            \hline
            Standing & 0.79 & 0.56 & 0.66 & 842\\
            Sitting & 0.67 & 0.95 & 0.78 & 827\\
            Walking & 0.74 & 0.66 & 0.70 & 1396\\
            Up/Downstairs & 0.48 & 0.58 & 0.53 & 676\\
            Stand-To-Sit & 0.80 & 0.14 & 0.23 & 59\\
            Sit-To-Stand & 0.60 & 0.14 & 0.22 & 66\\
            \hline
            \end{tabular}

    \label{tab:linear_probe_xformer_clf_report}
\end{table}

On SBHARPT, the convolutional encoder outperforms transformer-based encoder across both dynamic and static activities. It achieves better F1-scores across all activity classes, including very high score on lying down (F1 = 0.99) and strong results on walking upstairs, sitting and standing (around 0.86 each). The transformer-based encoder demonstrates solid performance in most activity classes except walking downstairs, achieving only F1-score of 0.63. Similar observations can be seen on FORTH-TRACE dataset. For dynamic and static activities, both encoders perform fairly well, but convolutional encoder is consistently better. For instance, it achieves F1-scores of 0.86 nd 0.84 for sitting and walking while transformer-based encoder drops to 0.53 on walking upstairs/downstairs. 

For transitional activities, performance drops for both encoders due to the small number of instances. This trend can be observed on both datasets. On SBHARPT, the convolutional encoder achieves better performance across all transitional activities, except lie-to-sit, achieving F1-scores ranging from 0.49 - 0.74, with stronger results on sit-to-stand and stand-to-sit and stand-to-lie activities. On FORTH-TRACE, the convolutional encoder achieves F1-scores of 0.40 and 0.38, indicating it still retains some predictive signals. In constrast, the transformer achieves very low F1-scores from 0.22 to 0.23, driven by extremely poor recall scores. 

The drop in classification performance for transitional activities is not only due to small number of instances. Unlike dynamic activities which contains repetitive motion cycles, and static activities, which exhibits relatively stable signal patterns, transitional activities are characterized by their brief duration that involve continuous changes in body posture. Consequently, only a limited portion of the window segmentation contains the sensor data corresponding to the actual transitional activity, while the remaining samples belong to the preceding or succeeding activities. As a result, the extracted features are less discriminative, and increase the likelihood of misclassification.

\begin{figure}[t]
    \centering
    \begin{subfigure}[b]{0.48\textwidth}
        \centering
        \includegraphics[scale=0.3]{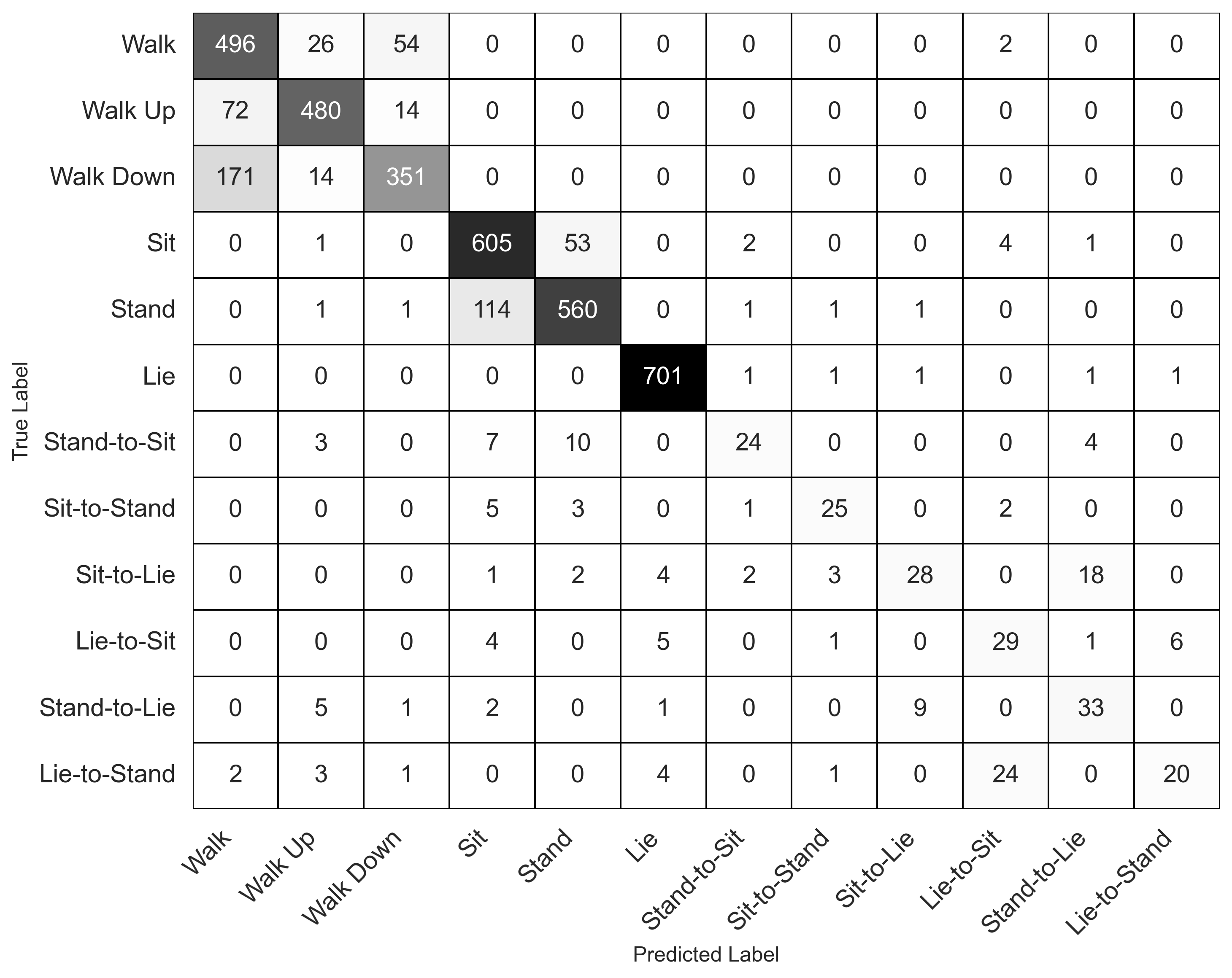}
        \caption{}
        \label{fig:fig2a_conf_matrix_conv_sbharpt}
    \end{subfigure}
    \hfill
    \begin{subfigure}[b]{0.48\textwidth}
        \centering
        \includegraphics[scale=0.3]{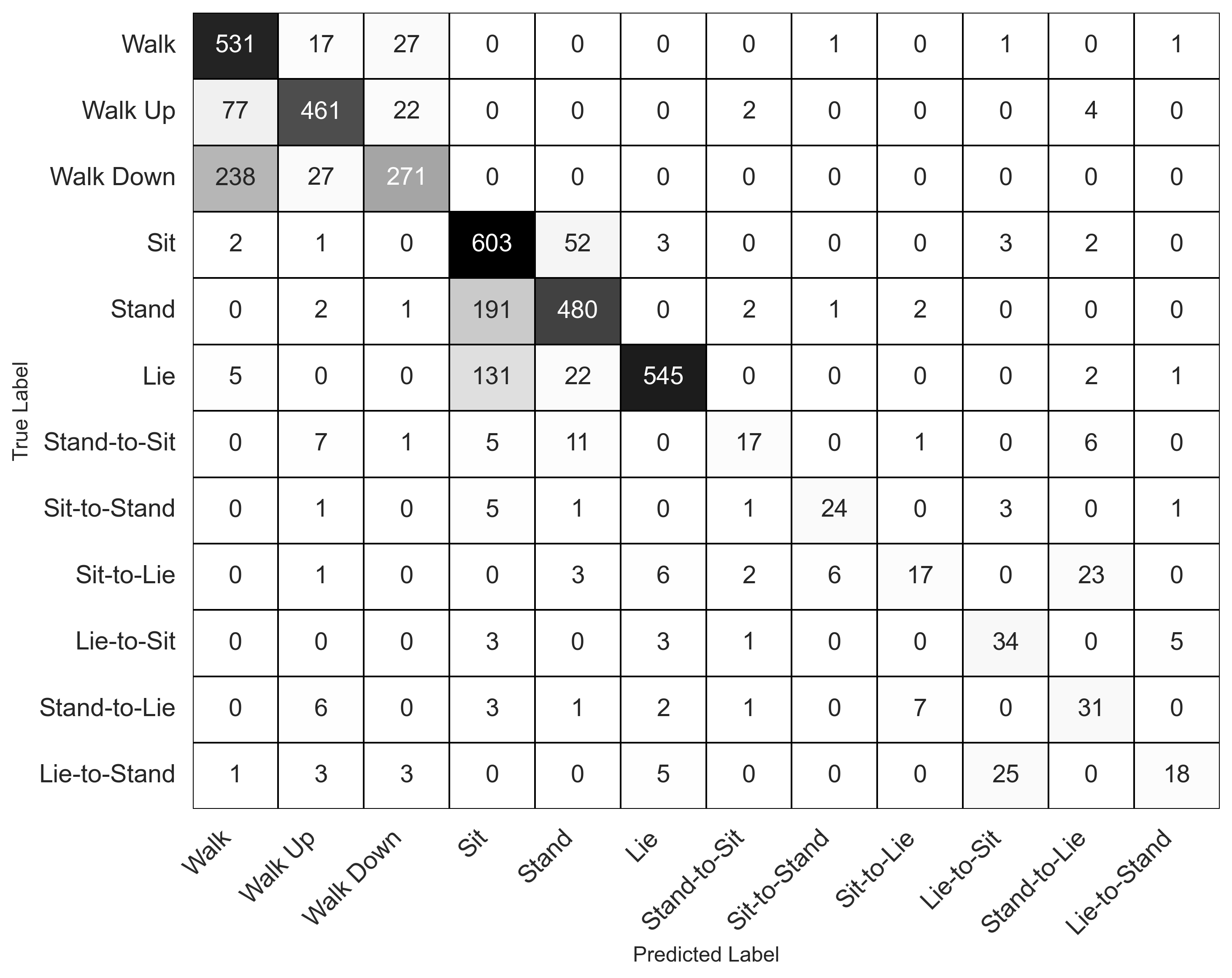}
        \caption{}
        \label{fig:fig2b_conf_matrix_xformer_sbharpt}
    \end{subfigure}
    \caption{Confusion matrix of the SBHARPT dataset for (a) Convolutional Encoder and (b) Transformer-based Encoder.}
    \label{fig:fig2_conf_matrices_sbharpt}
\end{figure}

\begin{figure}[t]
    \centering
    \begin{subfigure}[b]{0.48\textwidth}
        \centering
        \includegraphics[scale=0.3]{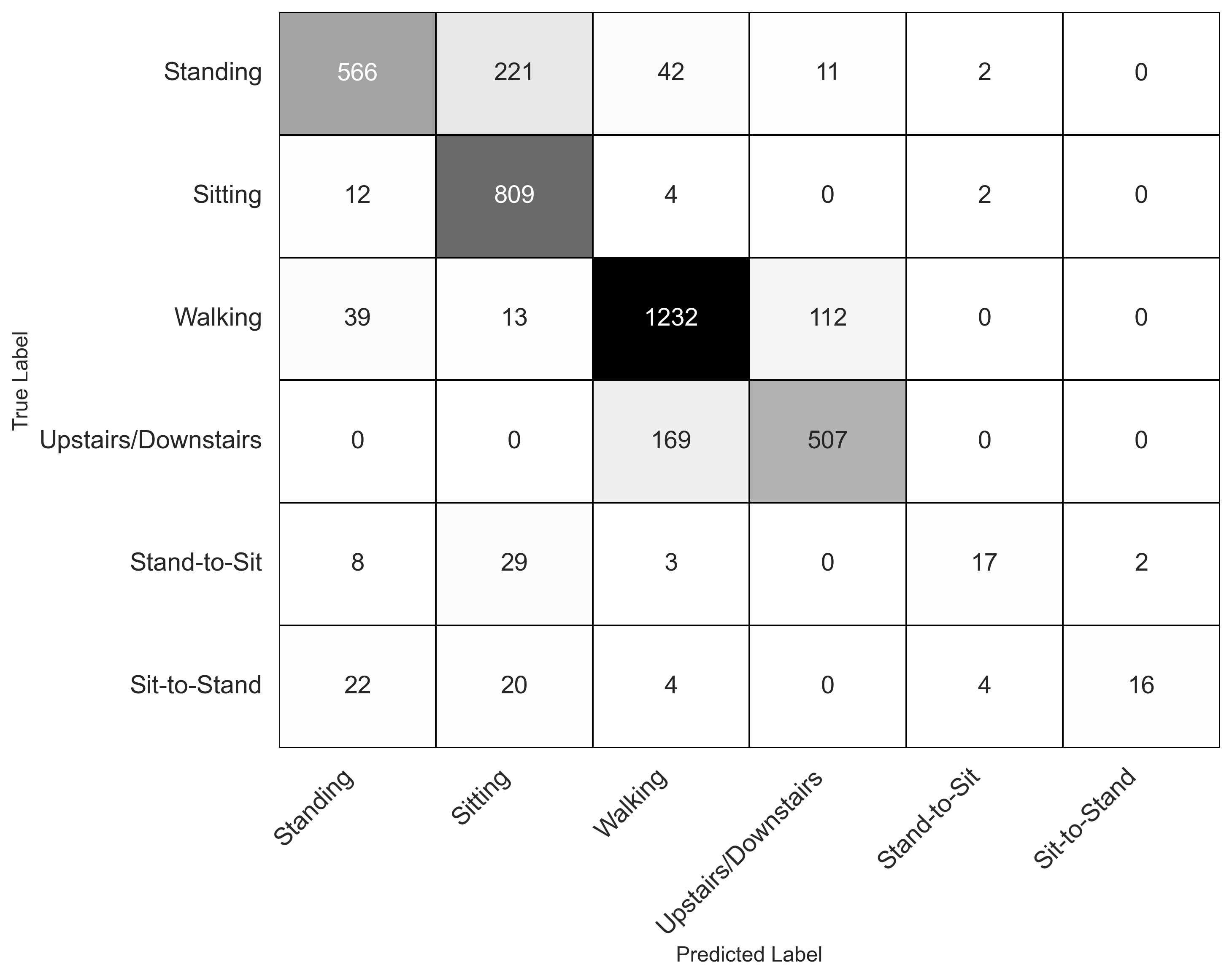}
        \caption{}
        \label{fig:fig3a_conf_matrix_conv_forthtrace}
    \end{subfigure}
    \hfill
    \begin{subfigure}[b]{0.48\textwidth}
        \centering
        \includegraphics[scale=0.3]{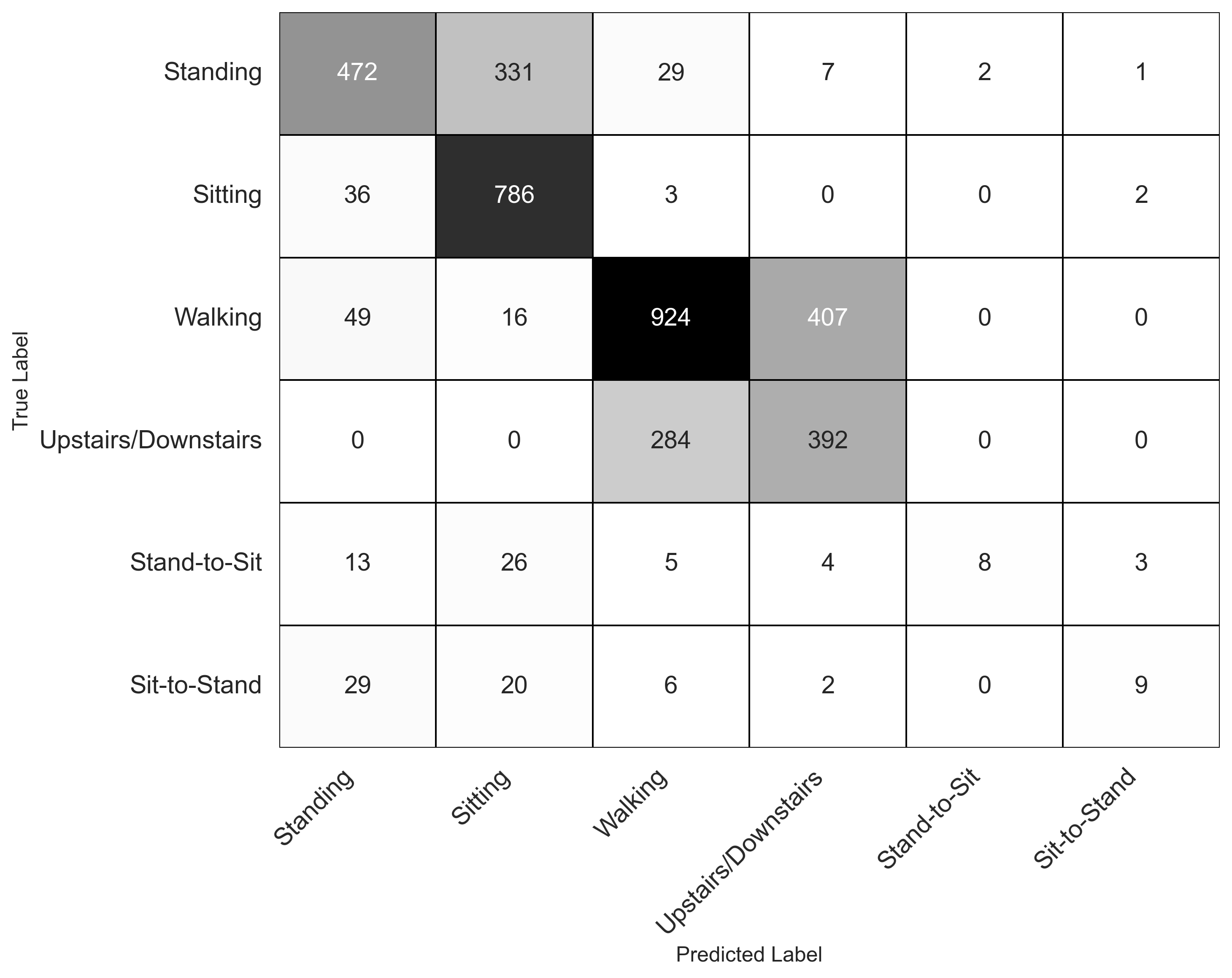}
        \caption{}
        \label{fig:fig3b_conf_matrix_xformer_forthtrace}
    \end{subfigure}
    \caption{Confusion matrix of the FORTH-TRACE dataset for (a) Convolutional Encoder and (b) Transformer-based Encoder.}
    \label{fig:fig3_conf_matrices_forthtrace}
\end{figure}

The confusion matrices of the linear probing across both datasets for both encoders are given in Figure \ref{fig:fig2_conf_matrices_sbharpt} and Figure \ref{fig:fig3_conf_matrices_forthtrace}. Analyzing the confusion matrices reveals that both encoders successfully classify activities that are different in motion patterns but struggle with similar activities. For instance, both architectures often confuse sitting and standing because the sensor signals of these two activities are very similar and are mainly influenced by body orientation due to gravity. Similarly, dynamic activities such as walking, walking upstairs and walking downstairs are also often cross-class misclassified due to their similar motion patterns. 

As for transitional activities, the confusion matrices show that the activities are more challenging to classify than dynamic and static activities. Analyzing the confusion matrices reveals that both encoders often confuse the transitional activities with their adjacent states such as stand-to-sit to standing or sitting related activities, and sit-to-lie to sitting or lying down related activities. Furthermore, sit-to-lie and lie-to-sit are often confused with stand-to-lie and lie-to-stand, respectively because the former transitions form part of the latter and therefore exhibit highly similar signal patterns.

The analysis suggests that the convolutional encoder is more effective in capturing localized temporal patterns associated with brief transitional activities, while transformer's ability to capture global temporal representations appears less effective in modeling these short-duration signal patterns. In contrast, both encoders demonstrate relatively strong performance on sustained activities such as walking, standing and sitting, whose signal patterns are characterized by more stable and discriminative features. These differences highlight the effectiveness of convolutional inductive biases such as temporal locality and translation invariance in feature extraction from multi-channel sensor window sequences with structural augmentations such as masking, jittering and channel reversal.

\subsection{Interpretation of the Latent Space}
\begin{figure}[b]
    \centering
    \begin{subfigure}{0.48\textwidth}
        \centering
        \includegraphics[scale=0.31]{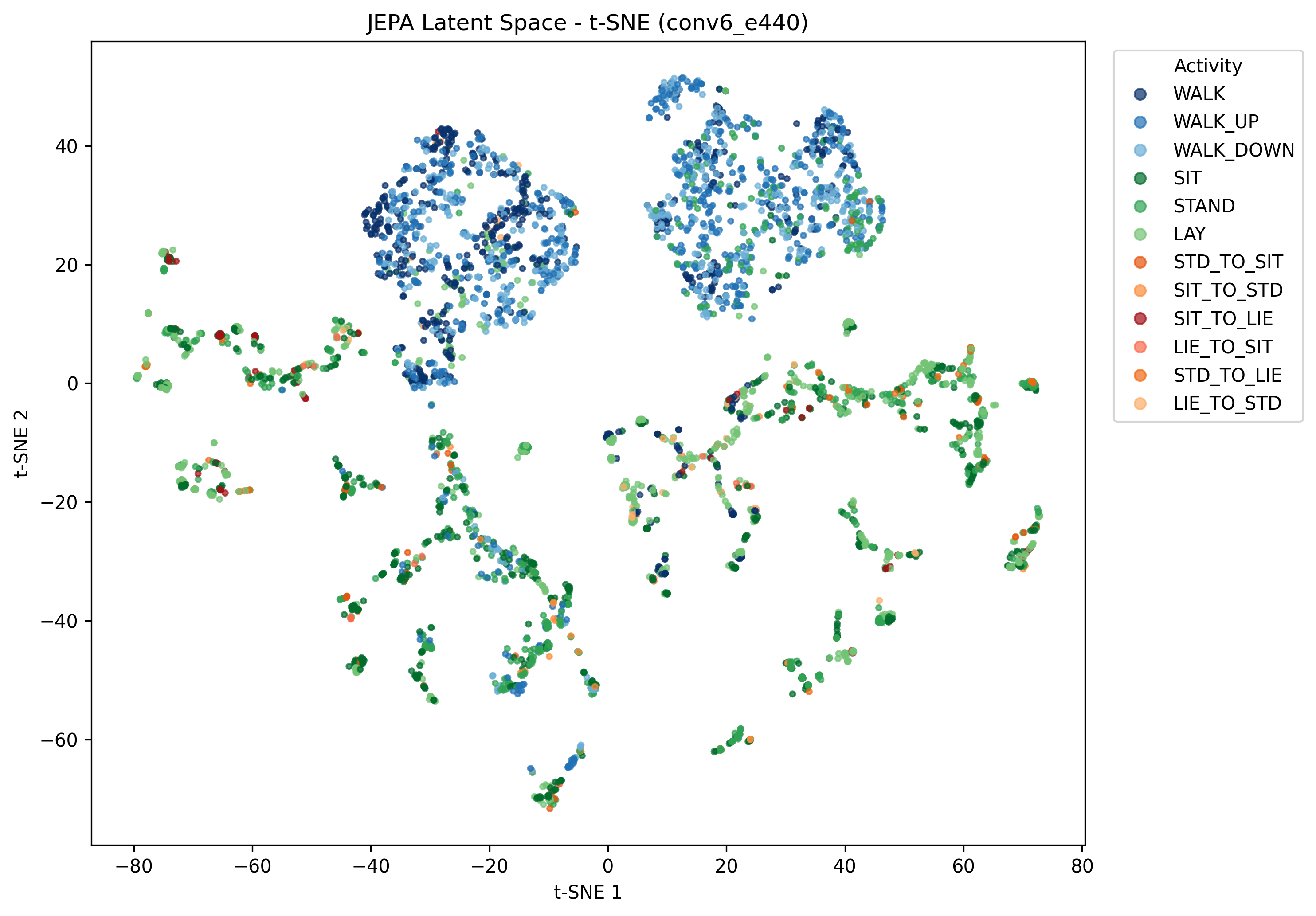}
        \caption{}
        \label{fig:jepa_latent_space_conv6_tnse_sbharpt}
    \end{subfigure}
    \hfill
    \begin{subfigure}{0.48\textwidth}
        \centering
        \includegraphics[scale=0.31]{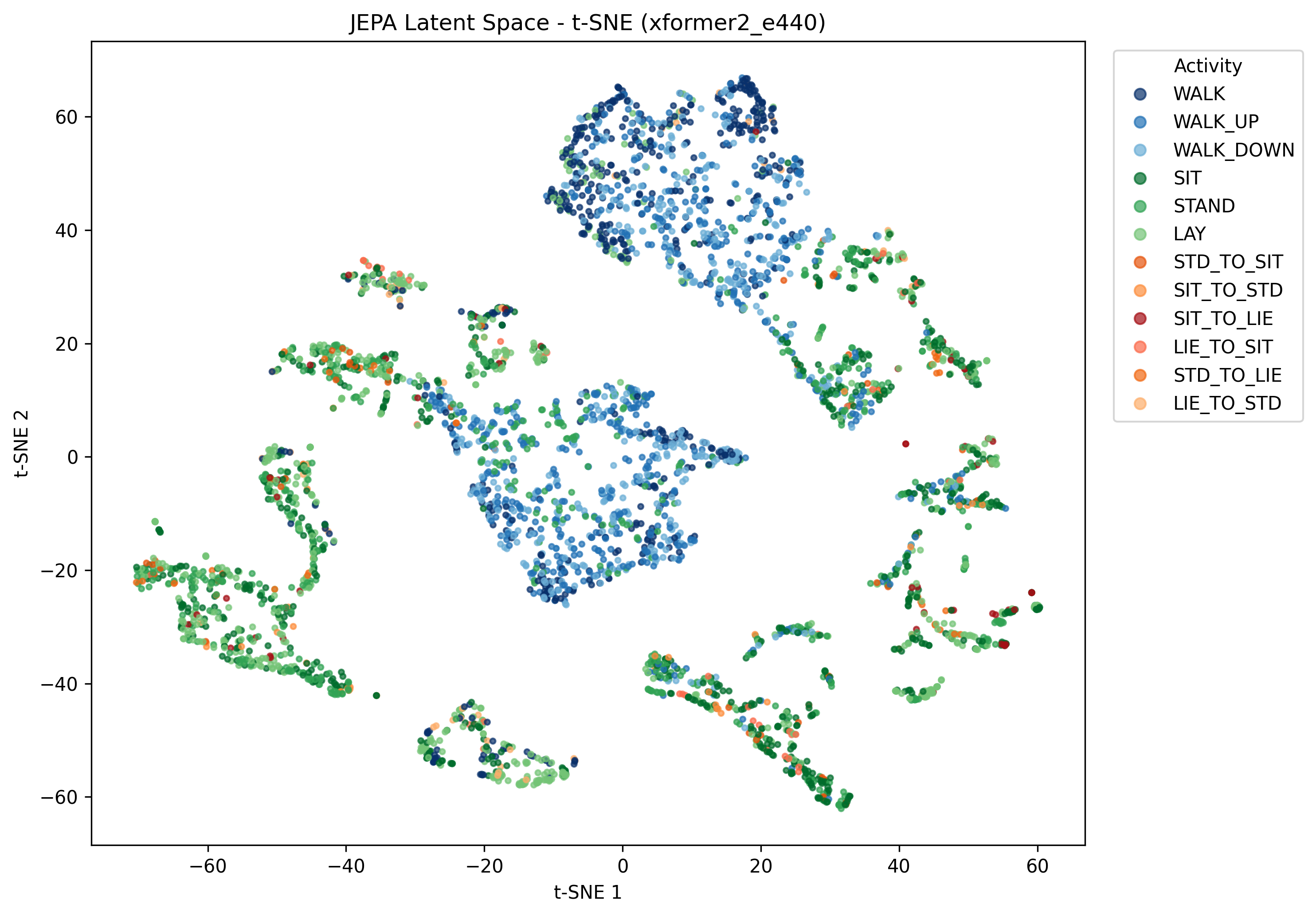}
        \caption{}
        \label{fig:jepa_latent_space_xformer2_tnse_sbharpt}
    \end{subfigure}
    \caption{Visualizations of learned embeddings using t-SNE for (a) convolutional encoder and (b) transformer-based encoder.}
    \label{fig:fig4_jepa_latent_space_tnse}
\end{figure}

The learned embeddings were qualitatively assessed using t-SNE manifold visualization. Figure \ref{fig:fig4_jepa_latent_space_tnse} presents the latent space for SBHARPT dataset. The visualizations show structural patterns that match the linear probing results, confirming that both encoders have learned meaningful feature spaces where similar activities are grouped together. This indicates that the proposed framework helps the JEPA models to organize semantically similar activities into nearby regions of the latent feature space.

As shown in the figure, the convolutional encoder produces a well-structured feature space, where the embeddings form two distinct, large clusters for dynamic activities, distinguishing walking on flat ground from walking on stairs. In the lower part of the feature space, the static and transitional activities are grouped into a large connected region. Although some transitional activities are partially overlap, similar activities are still placed close together. This organized representations allows a linear classifier to separate the classes easily, resulting in a superior linear probing performance. 

In contrast, the transformer-based encoder produces a more scattered feature space. The dynamic activities are still grouped together forming two distinct clusters near the top of the plot, but the clusters exhibit greater overlap than the convolutional ones. The static and transitional activities appear as long spread out patterns rather than compact clusters. This shows that transformer learns a more continuous representation of the sensor data instead of forming well-separated semantic activity clusters. As a result, the activity classes become less linearly separable, making them more difficult for a linear classifier to classify and leading to lower probing performance.

\subsection{Activity Classification}
The proposed framework is primarily designed to human activity recognition. This experiment evaluates the proposed framework by examining its classification effectiveness on the downstream task across both basic and transitional activities. Both encoder architectures are evaluated to determine how different inductive biases affect the classification.

\subsubsection{Convolutional Encoder} \label{eval_activity_conv_encoder}
After JEPA pre-training, the convolutional encoder is attached with a classification head. Then the classification model is fine-tuned using the progressive unfreezing training strategy as described in Section \ref{downstream_har}. The accuracy scores and F1-scores of the classification model for both datasets are presented in Table \ref{tab:conv_clf_performance}.

The results demonstrate that the proposed HAR-JEPA framework is effective for downstream HAR. On SBHARPT, the classification model achieves an accuracy of 0.9275 and an average F1-score of 0.8526, indicating a strong and balanced recognition across activity classes. For FORTH-TRACE, the classification model obtained an accuracy of 0.8702 and an average F1-score of 0.7763. Although the classification performance is lower than the performance on SBHARPT, the results remain competitive. This suggest that the learned embeddings are transferable to datasets with different sensing characteristics and activity classes.

\begin{table}[bt]
    \caption{HAR classification performance of the convolutional model across SBHARPT and FORTH-TRACE datasets}    
    \centering
    \small
    \begin{tabular}{ccc}
            \hline
            Dataset & Accuracy & Average F1-score \\  
            \hline
            SBHARPT & 0.9275 & 0.8526\\ 
            FORTH-TRACE & 0.8702 & 0.7763\\
            \hline
            \end{tabular}
    \label{tab:conv_clf_performance}
\end{table}

Table \ref{tab:conv_clf_report} presents the classification report for each activity for both datasets. Regarding dynamic and static activities, the classification model achieves robust performance compared to transitional activities, which remain considerably more difficult to classify across both datasets. For SBHARPT, walking achieves the highest F1-score of 0.99, closely followed by lying down at 0.98, while sitting and standing shared the lowest F1-score of 0.86. In FORTH-TRACE, walking and up/downstairs achieves the highest F1-score performance of 0.94, while standing had the lowest at 0.75.

\begin{table}[bt]
    \caption{HAR classification report of the convolutional model across SBHARPT and FORTH-TRACE datasets, reported in terms of precision, recall, F1-score, and support for each activity class.}    
    \centering
    \small
    \begin{tabular}{ccccc}
            \hline
            Activity & Precision & Recall & F1-score & Support\\  
            \hline
            &&SBHARPT \\
            \hline
            Walking & 0.98 & 0.99 & 0.99 & 578\\ 
            Walking Upstairs & 0.99 & 0.98 & 0.98 & 566\\ 
            Walking Downstairs & 0.98 & 0.99 & 0.98 & 536\\ 
            Sitting & 0.83 & 0.90 & 0.86 & 666\\ 
            Standing & 0.89 & 0.84 & 0.86 & 679\\ 
            Lying down & 0.98 & 0.98 & 0.98 & 706\\ 
            Stand-To-Sit & 0.92 & 0.73 & 0.81 & 48\\ 
            Sit-To-Stand & 0.87 & 0.92 & 0.89 & 36\\ 
            Sit-To-Lie & 0.80 & 0.67 & 0.73 & 58\\ 
            Lie-To-Sit & 0.66 & 0.76 & 0.71 & 46\\ 
            Stand-To-Lie & 0.76 & 0.75 & 0.75 & 51\\ 
            Lie-To-Stand & 0.77 & 0.60 & 0.67 & 55\\
            \hline
            &&FORTH-TRACE \\
            \hline
            Standing & 0.86 & 0.66 & 0.75 & 842\\
            Sitting & 0.75 & 0.96 & 0.84 & 827\\
            Walking & 0.92 & 0.96 & 0.94 & 1396\\
            Up/Downstairs & 0.96 & 0.91 & 0.94 & 676\\
            Stand-To-Sit & 0.82 & 0.53 & 0.64 & 59\\
            Sit-To-Stand & 0.84 & 0.41 & 0.55 & 66\\
            \hline
            \end{tabular}

    \label{tab:conv_clf_report}
\end{table}

\begin{figure}[t!]
    \centering
    \begin{subfigure}[b]{0.48\textwidth}
        \centering
        \includegraphics[scale=0.3]{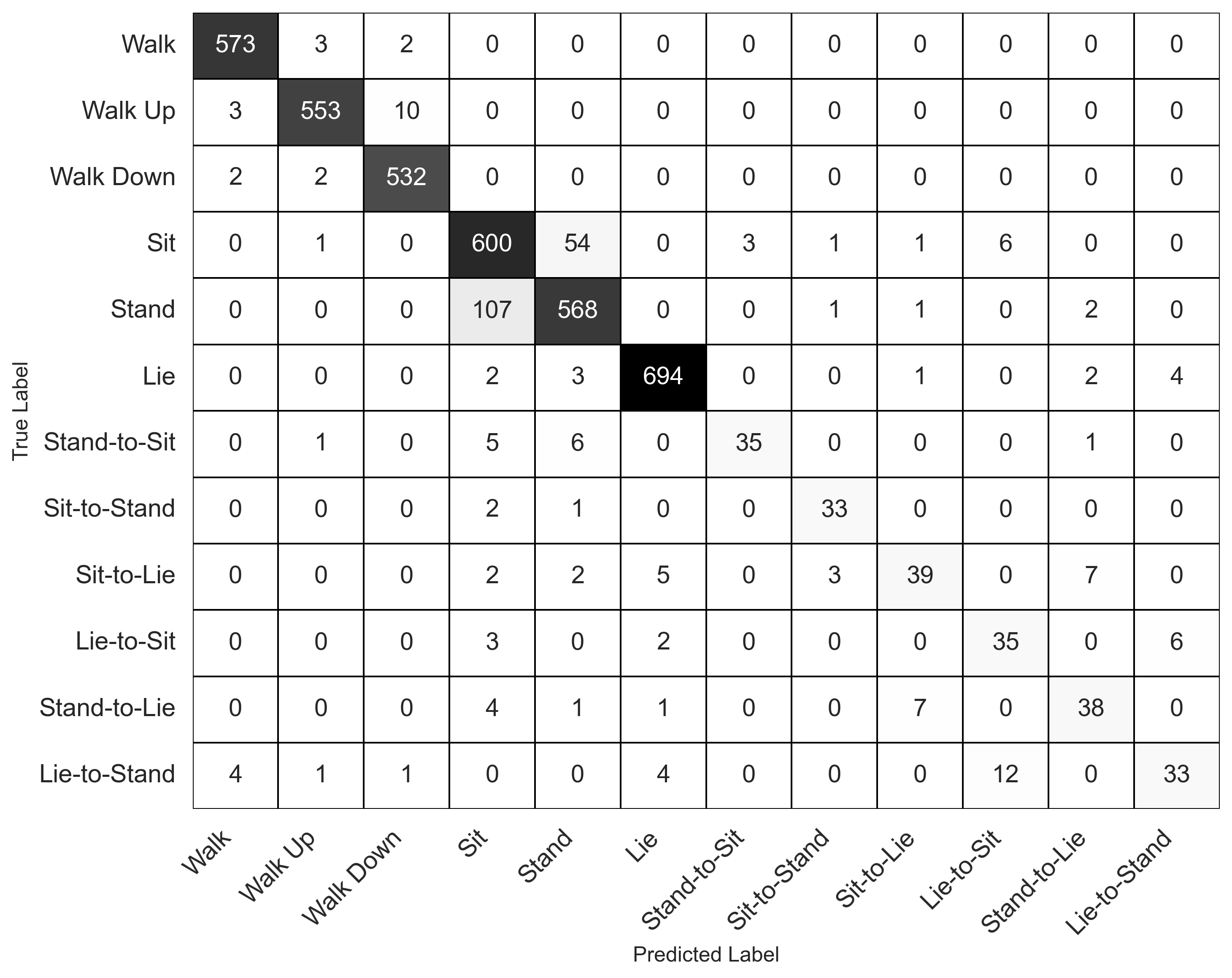}
        \caption{}
        \label{fig:fig6a_har_conf_matrix_conv_sbharpt}
    \end{subfigure}
    \hfill
    \begin{subfigure}[b]{0.48\textwidth}
        \centering
        \includegraphics[scale=0.3]{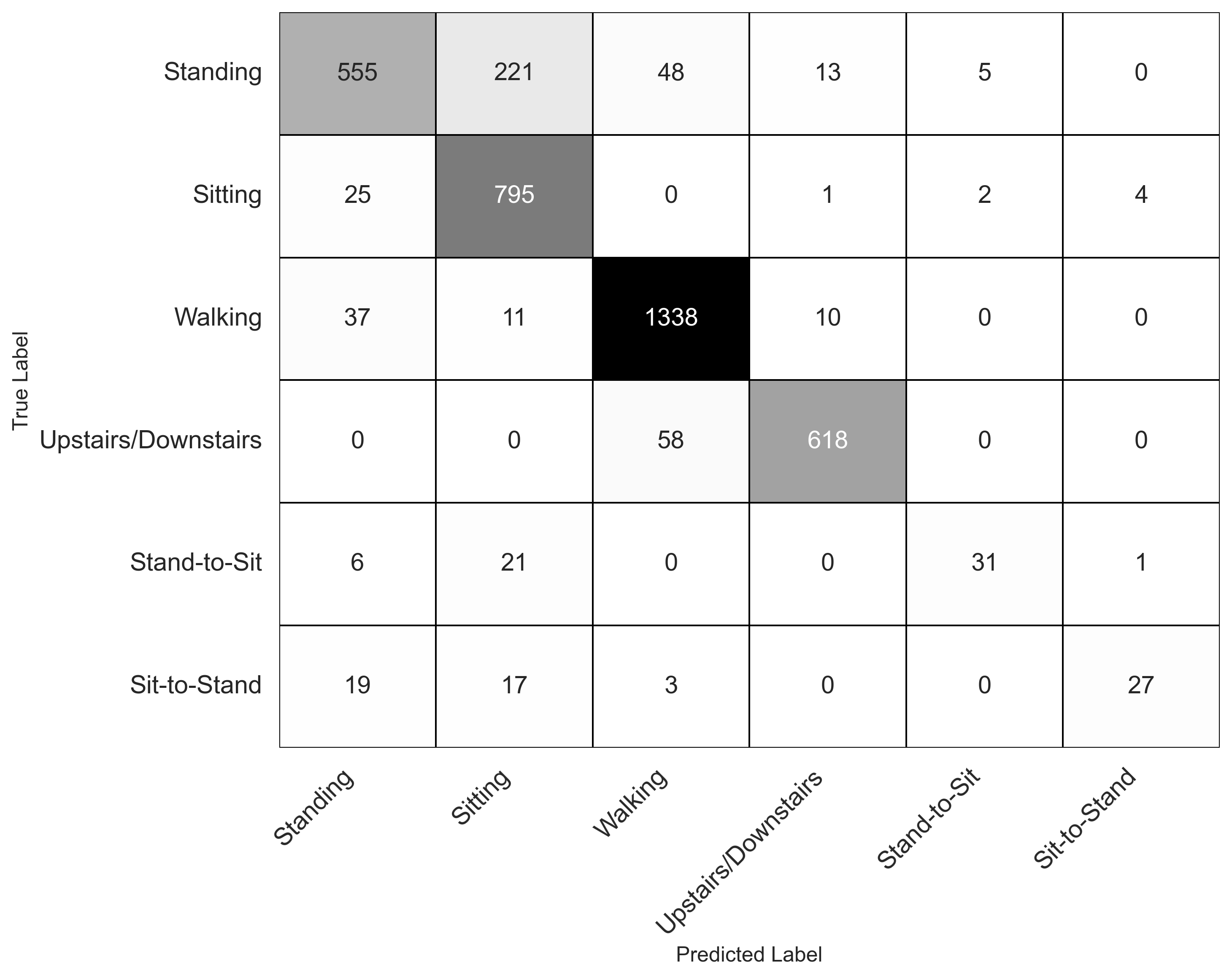}
        \caption{}
        \label{fig:fig6b_har_conf_matrix_conv_forthtrace}
    \end{subfigure}
    \caption{HAR confusion matrix of the convolutional model for (a) SBHARPT and (b) FORTH-TRACE.}
    \label{fig:fig6_conv_har_conf_matrices}
\end{figure}

As for transitional activities, sit-to-stand achieves the best F1-score performance in SBHARPT at 0.89, while in the FORTH-TRACE dataset, stand-to-sit achieves the highest F1-score at 0.64. Conversely, lie-to-stand achieves the lowest F1-score of 0.67 in the SBHARPT dataset, and sit-to-stand yielded the lowest F1-score of 0.55 in the FORTH-TRACE dataset.

These results are further supported by Figure \ref{fig:fig6_conv_har_conf_matrices}, which presents the confusion matrix of the classification model on the SBHARPT and FORTH-TRACE datasets. The confusion matrices reveal that the misclassification of dynamic activities mainly occur among the dynamic activity classes themselves. A similar pattern is observed for static activities, although a few instances are misclassified as transitional activities. By contrast, transitional activities exhibit substantially lower recognition rates, with frequent confusion with adjacent static states. 

The confusion matrix for the FORTH-TRACE dataset reveals that the misclassification of dynamic activities mainly occurs within the dynamic classes themselves, such as the overlap between walking and upstairs/downstairs. For static activities, misclassifications are mostly limited to other static states such as standing is frequently misclassified as sitting. By contrast, transitional activities show much lower recognition performance, with strong confusion with nearby static states. Specifically, stand-to-sit is most often misclassified as sitting, while sit-to-stand is commonly misclassified as standing and sitting. These results highlight the robustness of the model in recognizing stable, long-duration actions, while also highlighting the persistent challenge of accurately modeling short, context-dependent transitions.

\subsubsection{Transformer-based Encoder} \label{eval_activity_trans_encoder}

The transformer-based encoder is well-suited for capturing the broader context of a sequence. Unlike convolutional layers, which are constrained by the fixed size of their kernels and therefore focus only on local patterns, the transformer leverages the attention mechanism to evaluate the relative importance of every time step across the entire sequence. This mechanism enables it to model long-range dependencies, preserve global relationships, and discover subtle temporal dynamics that convolutional layers may overlook. 

The classification performance for both datasets is presented in Table \ref{tab:xformer_clf_performance}. Similar trends are observed, whereby the classification model demonstrates strong and consistent performance across both datasets, albeit slighly lower than that of the convolutional model. For SBHARPT, the classification model achieves an accuracy of 0.9292 and an average F1-score of 0.8588 while for FORTH-TRACE, it achieves an accuracy of 0.8632 and an average F1-score of 0.7402.

\begin{table}[bt]
    \caption{Classification Performance of the transformer-based model across SBHARPT and FORTH-TRACE datasets}    
    \centering
    \small
    \begin{tabular}{ccc}
            \hline
            Dataset & Accuracy & Average F1-score \\  
            \hline
            SBHARPT & 0.9292 & 0.8588\\ 
            FORTH-TRACE & 0.8632 & 0.7402\\
            \hline
            \end{tabular}
    \label{tab:xformer_clf_performance}
\end{table}

Table \ref{tab:xformer_clf_report} reports the model's classification report for each activity in terms of precision, recall, and F1-score. The table shows that dynamic and static activities consistently achieve higher F1-score, while transitional activities remain considerably more difficult to classify. Across both datasets, basic activities demonstrate robust performance. Lying down achieves consistently high F1-scores of 0.98 in the SBHARPT, while walking achieves the highest F1-score in the FORTH-TRACE dataset at 0.92. Dynamic activities including walking, walking upstairs and walking downstairs achieve very strong and consistent results in SBHARPT, with an F1-score of at least 0.96 across the three classes. Similarly, these activities in FORTH-TRACE achive an F1-score of at least 0.91 across the three classes.

\begin{table}[bt]
    \caption{HAR classification performance of the transformer-based model across SBHARPT and FORTH-TRACE datasets}    
    \centering
    \small
    \begin{tabular}{ccccc}
            \hline
            Activity & Precision & Recall & F1-score & Support\\  
            \hline
            &&SBHARPT \\
            \hline
            Walking & 0.96 & 0.99 & 0.97 & 578\\ 
            Walking Upstairs & 0.99 & 0.96 & 0.97 & 566\\ 
            Walking Downstairs & 0.96 & 0.97 & 0.96 & 536\\ 
            Sitting & 0.86 & 0.91 & 0.88 & 666\\ 
            Standing & 0.90 & 0.87 & 0.89 & 679\\ 
            Lying down & 0.99 & 0.97 & 0.98 & 706\\ 
            Stand-To-Sit & 0.85 & 0.69 & 0.76 & 48\\ 
            Sit-To-Stand & 0.84 & 0.89 & 0.86 & 36\\ 
            Sit-To-Lie & 0.84 & 0.71 & 0.77 & 58\\ 
            Lie-To-Sit & 0.67 & 0.78 & 0.72 & 46\\ 
            Stand-To-Lie & 0.77 & 0.78 & 0.78 & 51\\ 
            Lie-To-Stand & 0.81 & 0.71 & 0.76 & 55\\
            \hline
            &&FORTH-TRACE \\
            \hline
            Standing & 0.89 & 0.67 & 0.77 & 842\\
            Sitting & 0.76 & 0.98 & 0.85 & 827\\
            Walking & 0.90 & 0.95 & 0.92 & 1396\\
            Up/Downstairs & 0.95 & 0.87 & 0.91 & 676\\
            Stand-To-Sit & 0.67 & 0.49 & 0.57 & 59\\
            Sit-To-Stand & 0.95 & 0.27 & 0.42 & 66\\
            \hline

            \end{tabular}

    \label{tab:xformer_clf_report}
\end{table}

\begin{figure}[t!]
    \centering
    \begin{subfigure}[b]{0.48\textwidth}
        \centering
        \includegraphics[scale=0.3]{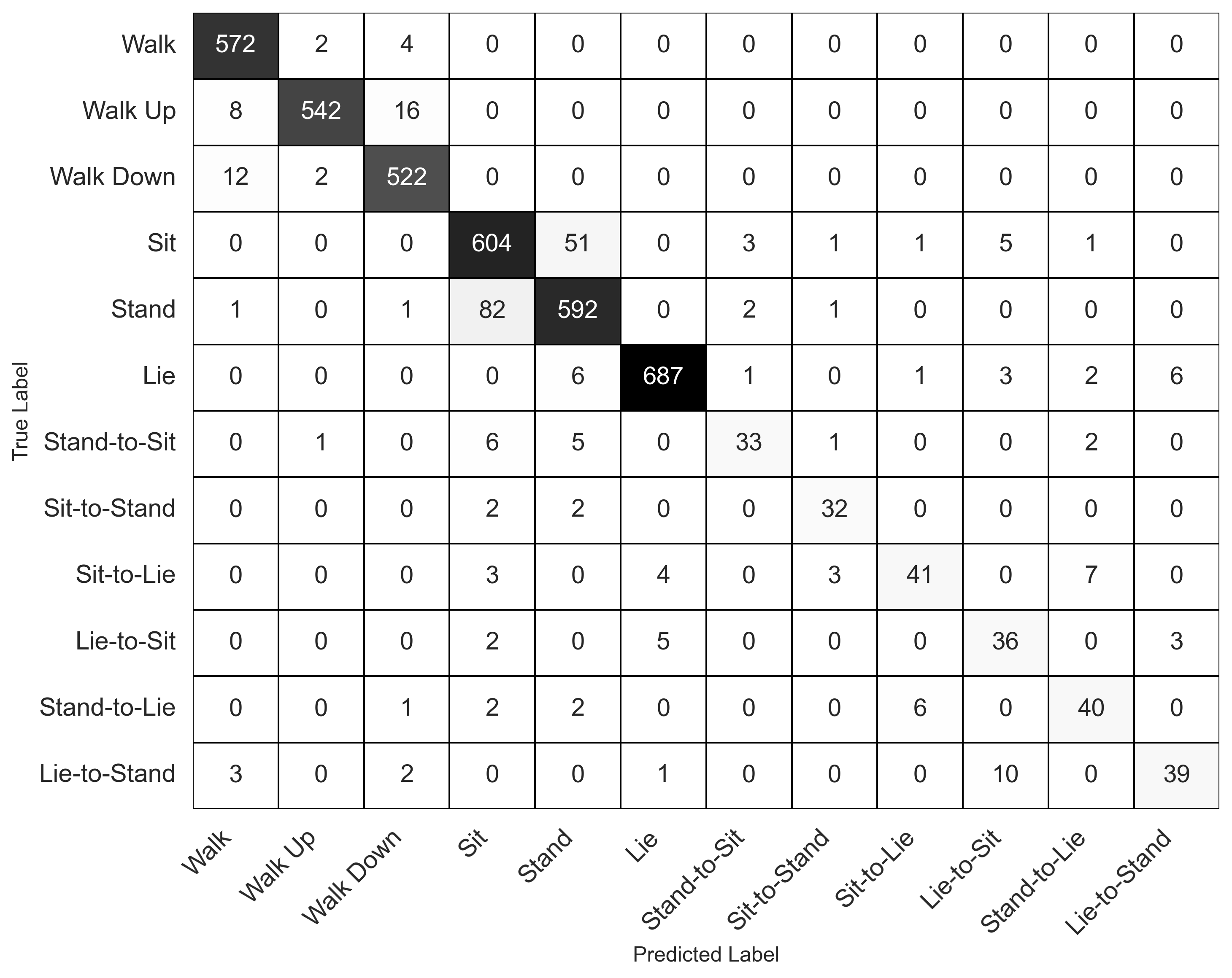}
        \caption{}
        \label{fig:fig5a_har_conf_matrix_xformer_sbharpt}
    \end{subfigure}
    \hfill
    \begin{subfigure}[b]{0.48\textwidth}
        \centering
        \includegraphics[scale=0.3]{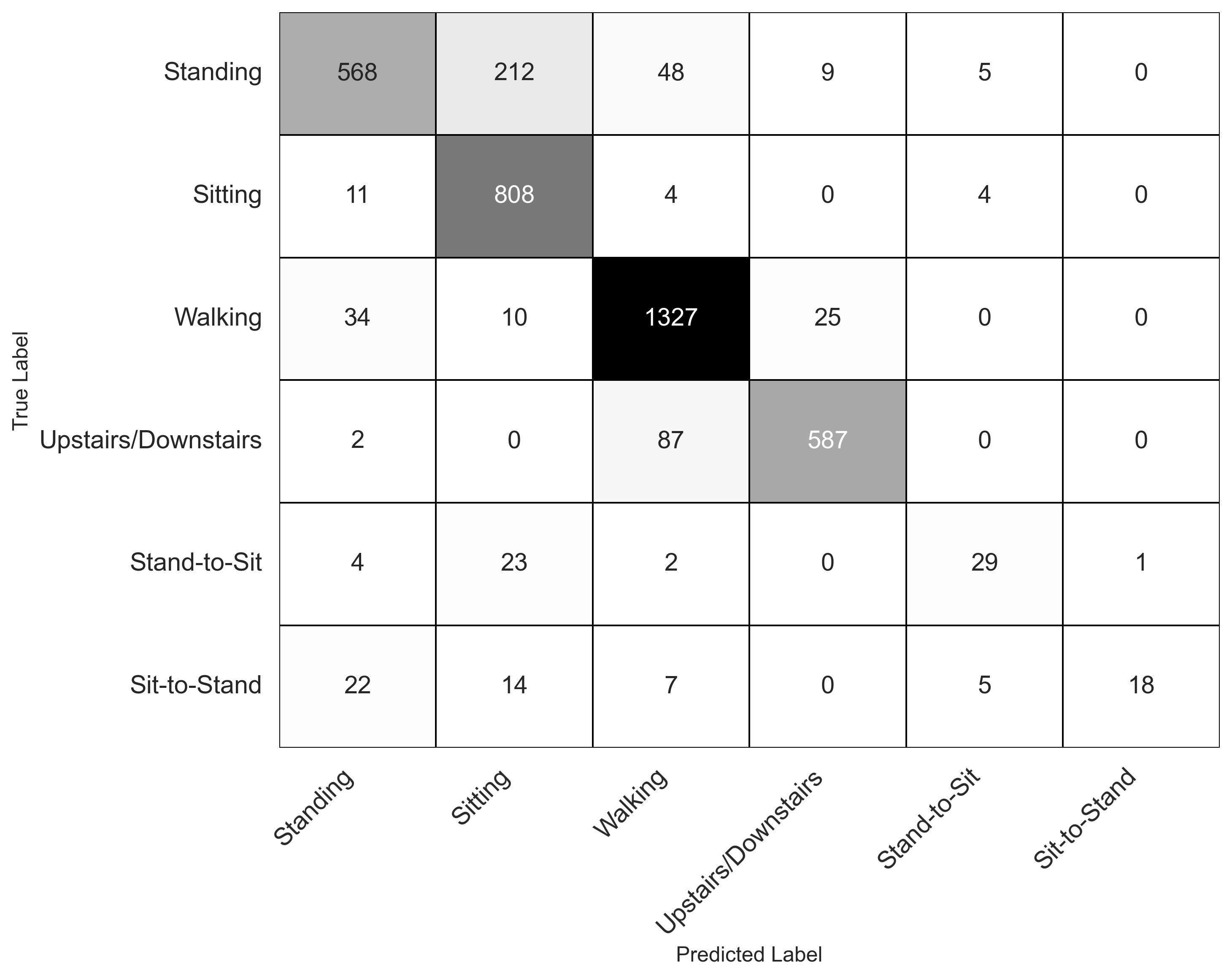}
        \caption{}
        \label{fig:fig5b_har_conf_matrix_xformer_forthtrace}
    \end{subfigure}
    \caption{HAR confusion matrix of the transformer-based model for (a) SBHARPT and (b) FORTH-TRACE.}
    \label{fig:fig5_xformer_har_conf_matrices}
\end{figure}

By contrast, transitional activities exhibit lower scores. While sit-to-stand in the SBHARPT dataset achieves an F1-score of 0.86, the other transitional activities achieve and an F1-score between 0.72 and 0.78. On FORTH-TRACE, stand-to-sit achieves an F1-score of 0.57, and sit-to-stand achieves an F1-score of 0.40, showing poor recognition. This discrepancy is largely explained by class imbalance and temporal signal characteristics whereby transitional activities occur less frequently, are shorter in duration, and often overlap with adjacent basic activities, which reduces the model’s ability to learn discriminative features.

The confusion matrices for both datasets are shown in Figure \ref{fig:fig5_xformer_har_conf_matrices}, providing further insight into the classification performance across dynamic, static and transitional activities. For the SBHARPT dataset, the confusion matrix reveals that misclassifications of dynamic activities occur mainly within the dynamic classes themselves, primarily involving the overlap of walking upstairs and walking downstairs with walking. Static activities follow a similar pattern of within-group confusion, which is mostly between sitting and standing whereby 82 instances of standing are misclassified as sitting, and 51 instances of sitting are misclassified as standing. By contrast, transitional activities exhibit significantly lower recognition rates and are frequently confused with adjacent static states or other transitions such as stand-to-lie is often misclassified as sit-to-lie, and lie-to-stand is misclassified as lie-to-sit and walking.

The confusion matrix for the FORTH-TRACE similarly reveals that misclassification of dynamic activities mainly occurs within the dynamic activity classes, characterized by a substantial overlap where up/downstairs is misclassified as walking. For static activities, asymmetrical misclassification occurs from standing to sitting whereas sitting is rarely misclassified as standing. As for transitional activities, the matrix shows much weaker recognition performance due to heavy confusion with their adjacent static states. Specifically, stand-to-sit is frequently misclassified as sitting, while sit-to-stand is commonly misclassified as standing and sitting.

The comparative analysis of SBHARPT and FORTH-TRACE datasets highlights consistent trends in the recognition of basic (dynamic and static) versus transitional activities. In both cases, basic activities such as walking, upstairs/downstairs, sitting, standing, and lying down achieve strong recognition, as reflected by high counts along the diagonal of the confusion matrices and limited misclassifications. 

By contrast, transitional activities remain a clear bottleneck. In addition, stand-to-sit and sit-to-stand are frequently misclassified as sitting or standing. More complex transitions, such as sit-to-lie and lie-to-stand, show substantial confusion, often being misclassified into adjacent stable states (lying or sitting). These results confirm that transitional activities, being shorter in duration and underrepresented in the dataset, are more prone to misclassification due to temporal overlap and feature similarity.

The results also highlight two critical points. First, accuracy alone is insufficient for evaluating HAR under imbalanced conditions, as it can obscure poor performance on minority classes. Metrics such as precision, recall, and F1-score provide a more balanced assessment by emphasizing minority class recognition. Second, the difficulty of modeling transitional activities highlights the need for specialized strategies, including segmentation methods, oversampling, or more complex architectures to better capture transitions.

\subsection{Effect of Dispersion Coefficient}
The proposed norm term of VICReg objective prevents the embeddings from collapsing into a single point at the origin. The target norm is governed by the dispersion coefficient $\sigma$ which determines the expansion of the target norm relative to the unit sphere. This section examines the effect of the coefficient on representation quality and downstream task performance. Table \ref{tab:sigma_accuracy_f1score} compare the performance of both encoders across different $\sigma$ values.

As shown in the table, the selection of dispersion coefficient $\sigma$ influence the HAR performance. For the SBHARPT, increasing the value yields performance gains across both architectures. Specifically, the average F1-score improves from 0.8407 to 0.8654 for convolutional encoder and from 0.8467 to 0.8621 for transformer-based encoder. A similar trend is observed for accuracy measure. For the FORTH-TRACE dataset, a contrasting trend is observed where larger $\sigma$ values degrade the accurate rate of the encoders. The accuracy drops from 0.8702 to 0.8652 for convolutional encoder while for transformer, the accuracy reduces slightly to 0.8572 from 0.8606. As for average F1-score, a similar trend is visible for the convolutional encoder, which peaks at 0.7763 before dropping to 0.7669. Interestingly, the transformer-based encoder experiences a drop at $\sigma=0.25$ and finally recovers to achieve its highest score of 0.7550 at $\sigma=0.30$. In summary, the choice of $\sigma$ and encoder architecture yields distinct performance trends, depending on the dataset. There is no single value of $\sigma$ that can be considered optimal. Lower $\sigma$ results in better-performing learned embeddings for FORTH-TRACE whereas a larger value produces better performance for SBHARPT. In both scenarios, the convolutional encoder demonstrates superior performance.

\begin{table}[bt]
\centering
\small
\caption{Performance comparison between both encoder architectures for each $\sigma$ values}
\label{tab:sigma_accuracy_f1score}
\begin{tabular}{lllcc}
\toprule
Dataset & Architecture & $\sigma$ & Accuracy & Average F1-score \\
\midrule
\multirow{6}{*}{SBHARPT} & \multirow{3}{*}{Conv} & 0.20 & 0.9277 & 0.8407 \\
 & & 0.25 & 0.9275 & 0.8526 \\
 & & 0.30 & 0.9337 & 0.8654 \\
\cmidrule{2-5}
 & \multirow{3}{*}{Transformer} & 0.20 & 0.9272 & 0.8467 \\
 & & 0.25 & 0.9292 & 0.8588 \\
 & & 0.30 & 0.9329 & 0.8621 \\
\midrule
\multirow{6}{*}{FORTH-TRACE} & \multirow{3}{*}{Conv} & 0.20 & 0.8702 & 0.7687 \\
 & & 0.25 & 0.8702 & 0.7763 \\
 & & 0.30 & 0.8652 & 0.7669 \\
\cmidrule{2-5}
 & \multirow{3}{*}{Transformer} & 0.20 & 0.8606 & 0.7498 \\
 & & 0.25 & 0.8632 & 0.7402 \\
 & & 0.30 & 0.8572 & 0.7550 \\
\bottomrule
\end{tabular}
\end{table}

\subsection{Comparison with Supervised Learning Models}
To ensure a fair comparison, supervised learning models are implemented using the same encoder architectures as HAR-JEPA, differing only in the training strategy. The results of each encoder are presented in Table \ref{tab:supervised_clf_performance}. Overall, the performance of the supervised convolutional model exceeds the transformer-based model for both datasets, achieving an average F1-score of 0.8663 and 0.7445 for SBHARPT and FORTH-TRACE respectively.

\begin{table}[bt]
    \caption{Accuracy and F1-score of the supervised models across SBHARPT and FORTH-TRACE datasets}    
    \centering
    \small
    \begin{tabular}{ccccc}
            \hline
            Encoder & Accuracy & F1-score \\           
            \hline
            &SBHARPT \\           
            \hline
            Convolutional & 0.9366 & 0.8663 \\ 
            Transformer-based & 0.9267 & 0.8482 \\         
            \hline
            &FORTH-TRACE \\  
            \hline
            Convolutional & 0.8577 & 0.7445 \\
            Transformer-based & 0.8430 & 0.7436 \\ 
            \hline
            \end{tabular}

    \label{tab:supervised_clf_performance}
\end{table}

The classification report for both supervised models are given in Tables \ref{tab:supervised_conv_clf_report} and \ref{tab:supervised_xformer_clf_report}. The supervised learning models achieve robust performance for dynamic and static activities compared to transitional activities, which remain considerably more difficult to classify due to their signal characteristics and low number of instances. For SBHARPT, walking and lying down consistently achieves the highest F1-score of above 0.98 across both models while sitting and standing achieve considerably lower scores in the range of 0.87 to 0.89.	 As for transitional activities, both models achieve the highest F1-score (above 0.83) in classifying stand-to-sit and sit-to-stand achieve. However, both models consistently struggle to identify sit-to-lie and stand-to-lie, achieving an F1-score in the range of 0.65 to 0.73. Similar trend is observed for FORTH-TRACE dataset in which both dynamic and static activities achieve better performance than transitional activities. Walking and walking upstairs/downstairs achieve the highest F1-scores, ranging from 0.89 to 0.93 across both models. For static activities, sitting achieves an F1-score of 0.84 to 0.85, whereas standing shows a performance drop to 0.72 (convolutional) and 0.73 (transformer).

\begin{table}[bt]
    \caption{HAR classification performance of the supervised convolutional model across SBHARPT and FORTH-TRACE datasets}    
    \centering
    \small
    \begin{tabular}{ccccc}
            \hline
            Activity & Precision & Recall & F1-score & Support\\  
            \hline
            &&SBHARPT \\
            \hline
            Walking & 0.98 & 0.99 & 0.99 & 578\\ 
            Walking Upstairs & 0.98 & 0.98 & 0.98 & 566\\ 
            Walking Downstairs & 0.97 & 0.99 & 0.98 & 536\\ 
            Sitting & 0.87 & 0.90 & 0.89 & 666\\ 
            Standing & 0.90 & 0.88 & 0.89 & 679\\ 
            Lying down & 0.98 & 0.99 & 0.99 & 706\\ 
            Stand-To-Sit & 0.90 & 0.77 & 0.83 & 48\\ 
            Sit-To-Stand & 0.85 & 0.81 & 0.83 & 36\\ 
            Sit-To-Lie & 0.74 & 0.69 & 0.71 & 58\\ 
            Lie-To-Sit & 0.82 & 0.78 & 0.80 & 46\\ 
            Stand-To-Lie & 0.75 & 0.71 & 0.73 & 51\\ 
            Lie-To-Stand & 0.89 & 0.71 & 0.79 & 55\\
            \hline
            &&FORTH-TRACE \\
            \hline
            Standing & 0.84 & 0.63 & 0.72 & 842\\
            Sitting & 0.77 & 0.94 & 0.85 & 827\\
            Walking & 0.89 & 0.96 & 0.92 & 1396\\
            Up/Downstairs & 0.96 & 0.91 & 0.93 & 676\\
            Stand-To-Sit & 0.83 & 0.42 & 0.56 & 59\\
            Sit-To-Stand & 0.79 & 0.35 & 0.48 & 66\\
            \hline
            \end{tabular}

    \label{tab:supervised_conv_clf_report}
\end{table}

\begin{table}[bt!]
    \caption{HAR classification performance of the supervised transformer-based model across SBHARPT and FORTH-TRACE datasets}    
    \centering
    \small
    \begin{tabular}{ccccc}
            \hline
            Activity & Precision & Recall & F1-score & Support\\  
            \hline
            &&SBHARPT \\
            \hline
            Walking & 0.96 & 0.99 & 0.98 & 578\\ 
            Walking Upstairs & 0.99 & 0.96 & 0.97 & 566\\ 
            Walking Downstairs & 0.96 & 0.97 & 0.97 & 536\\ 
            Sitting & 0.85 & 0.90 & 0.87 & 666\\ 
            Standing & 0.91 & 0.86 & 0.88 & 679\\ 
            Lying down & 0.98 & 0.99 & 0.98 & 706\\ 
            Stand-To-Sit & 0.85 & 0.83 & 0.84 & 48\\ 
            Sit-To-Stand & 0.84 & 0.89 & 0.86 & 36\\ 
            Sit-To-Lie & 0.70 & 0.66 & 0.68 & 58\\ 
            Lie-To-Sit & 0.70 & 0.72 & 0.71 & 46\\ 
            Stand-To-Lie & 0.62 & 0.69 & 0.65 & 51\\ 
            Lie-To-Stand & 0.90 & 0.67 & 0.77 & 55\\ 
            \hline
            &&FORTH-TRACE \\
            \hline
            Standing & 0.89 & 0.62 & 0.73 & 842\\
            Sitting & 0.74 & 0.97 & 0.84 & 827\\
            Walking & 0.90 & 0.91 & 0.90 & 1396\\
            Up/Downstairs & 0.87 & 0.90 & 0.89 & 676\\
            Stand-To-Sit & 0.71 & 0.51 & 0.59 & 59\\
            Sit-To-Stand & 0.64 & 0.42 & 0.51 & 66\\
            \hline
            \end{tabular}

    \label{tab:supervised_xformer_clf_report}
\end{table}

Comparing to the best performing HAR-JEPA models, in general, the proposed models achieve a better performance in terms of F1-score. Tables \ref{tab:JEPA_best_conv_clf_report} and \ref{tab:JEPA_best_xformer_clf_report} show the classification reports of the HAR-JEPA-based models. For the SBHARPT dataset, the HAR-JEPA convolutional model demonstrates competitive results against its supervised counterpart, achieving near-identical F1-scores above 0.98 for dynamic and static activities. Crucially, the HAR-JEPA models exhibit a noticeable advantage in classifying the more challenging transitional activities achieving F1-score, ranging from 0.75 to 0.85 across transition classes. This shows that the JEPA pre-training objective learn more discriminative temporal representations for activities with complex motion patterns and limited training instances. Similar trend is observed for HAR-JEPA transformer-based model in which it achieves comparable performance for dynamic and static activities, while improving the F1-score (0.70 to 0.87) of transitional activities relative to its supervised counterpart (0.65 to 0.86).

This performance improvement is also observed on the FORTH-TRACE dataset. For the HAR-JEPA convolutional model, all activity classes achieve better F1-scores compared to its supervised counterpart, with the most noticeable improvement observed for the transitional activities, where the scores for stand-to-sit and sit-to-stand increase by 0.06 and 0.04, respectively. The HAR-JEPA transformer-based model demonstrates similar trend, improving the performance of most activity classes except sit-to-stand where the F1-score drops by 0.03. Overall the results indicate that HAR-JEPA pre-training framework enhances feature representation, leading to better generalization across dynamic, static and challenging transitional activities.

\begin{table}[bt]
    \caption{HAR classification performance of the HAR-JEPA convolutional models across SBHARPT ($\sigma=0.30$) and FORTH-TRACE ($\sigma=0.20$) datasets}    
    \centering
    \small
    \begin{tabular}{ccccc}
            \hline
            Activity & Precision & Recall & F1-score & Support\\  
            \hline
            &&SBHARPT \\
            \hline
            Walking & 0.98 & 0.99 & 0.99 & 578\\ 
            Walking Upstairs & 0.98 & 0.97 & 0.98 & 566\\ 
            Walking Downstairs & 0.97 & 0.98 & 0.98 & 536\\ 
            Sitting & 0.86 & 0.90 & 0.88 & 666\\ 
            Standing & 0.89 & 0.88 & 0.88 & 679\\ 
            Lying down & 0.98 & 0.98 & 0.98 & 706\\ 
            Stand-To-Sit & 0.94 & 0.71 & 0.81 & 48\\ 
            Sit-To-Stand & 0.82 & 0.89 & 0.85 & 36\\ 
            Sit-To-Lie & 0.78 & 0.74 & 0.76 & 58\\ 
            Lie-To-Sit & 0.76 & 0.74 & 0.75 & 46\\ 
            Stand-To-Lie & 0.84 & 0.73 & 0.78 & 51\\ 
            Lie-To-Stand & 0.84 & 0.67 & 0.75 & 55\\
            \hline
            &&FORTH-TRACE \\
            \hline
            Standing & 0.89 & 0.64 & 0.74 & 842\\
            Sitting & 0.75 & 0.97 & 0.85 & 827\\
            Walking & 0.91 & 0.96 & 0.94 & 1396\\
            Up/Downstairs & 0.97 & 0.92 & 0.94 & 676\\
            Stand-To-Sit & 0.85 & 0.49 & 0.62 & 59\\
            Sit-To-Stand & 0.76 & 0.39 & 0.52 & 66\\
            \hline
            \end{tabular}

    \label{tab:JEPA_best_conv_clf_report}
\end{table}

\begin{table}[bt!]
    \caption{HAR classification performance of the HAR-JEPA transformer-based models across SBHARPT and FORTH-TRACE datasets}    
    \centering
    \small
    \begin{tabular}{ccccc}
            \hline
            Activity & Precision & Recall & F1-score & Support\\  
            \hline
            &&SBHARPT \\
            \hline
            Walking & 0.96 & 0.99 & 0.97 & 578\\ 
            Walking Upstairs & 0.99 & 0.95 & 0.97 & 566\\ 
            Walking Downstairs & 0.96 & 0.98 & 0.97 & 536\\ 
            Sitting & 0.89 & 0.89 & 0.89 & 666\\ 
            Standing & 0.90 & 0.90 & 0.90 & 679\\ 
            Lying down & 0.99 & 0.98 & 0.98 & 706\\ 
            Stand-To-Sit & 0.82 & 0.85 & 0.84 & 48\\ 
            Sit-To-Stand & 0.91 & 0.83 & 0.87 & 36\\ 
            Sit-To-Lie & 0.76 & 0.76 & 0.76 & 58\\ 
            Lie-To-Sit & 0.65 & 0.76 & 0.70 & 46\\ 
            Stand-To-Lie & 0.77 & 0.78 & 0.78 & 51\\ 
            Lie-To-Stand & 0.80 & 0.65 & 0.72 & 55\\
            \hline
            &&FORTH-TRACE \\
            \hline
            Standing & 0.86 & 0.66 & 0.75 & 842\\
            Sitting & 0.77 & 0.95 & 0.85 & 827\\
            Walking & 0.87 & 0.96 & 0.91 & 1396\\
            Up/Downstairs & 0.98 & 0.85 & 0.91 & 676\\
            Stand-To-Sit & 0.70 & 0.56 & 0.62 & 59\\
            Sit-To-Stand & 0.79 & 0.35 & 0.48 & 66\\
            \hline
            \end{tabular}

    \label{tab:JEPA_best_xformer_clf_report}
\end{table}

\subsection{Comparison with State-of-the-art Approaches}
We compare the proposed method with state-of-the-art approaches to HAR including fully supervised and self-supervised learning. However, it should be noted that the field currently lacks a universally accepted evaluation benchmark as variations in data preprocessing, segmentation (window) size, subject-independent versus subject-dependent splitting and cross-dataset evaluation protocols. For fully supervised learning, we focus on studies that utilize the same datasets while for self-supervised learning, we select representative contrastive learning-based studies as a point of comparison. Tables \ref{tab:compare_fully_supervised} and \ref{tab:compare_self_supervised} presents the comparison with fully supervised and self-supervised learning approaches respectively. 

In general, the proposed method achieves performance that is comparable to, if not better than existing fully supervised approaches reported in \cite{MohdNoor2021, Baraka2023, Baraka2024} on the same datasets. However, these studies employ adaptive sliding window and multiple predictive models to determine effective window size and distinguish between basic and transitional activity windows prior to classification. This approach introduces additional computational complexity and increases design complexity, which may limit its practicality for real-time or resource-constrained applications. 

Studies \cite{Lone2021, Irfan2021} achieve a higher accuracy on the same dataset. However, the study conducts its experiments by separating basic activity data from transitional activity data and evaluating them independently \cite{Lone2021}, while \cite{Irfan2021} builds multiple models and combines their predictions using ensemble learning. Another study \cite{Ige2025} does not use the same datasets, instead evaluating on PAMAP and WISDM, and employs a hybrid CNN-BiLSTM model with an attention mechanism to extract both local and temporal features, which are then combined into a single representation for activity classification.

\begin{sidewaystable}[p]
    \caption{Comparison between the proposed and supervised learning methods}    
        \centering
        \footnotesize
    \renewcommand{\arraystretch}{1.3}
    \setlength{\tabcolsep}{4pt}
    \begin{tabularx}{\textwidth}{X X X X}
            \hline
            Paper (Year) & Dataset (Metric) & Model & Description\\   
            \hline
            Lone et al. \cite{Lone2021} (2024) & SBHARPT (Acc): 0.9827 (Set 1), 0.9584 (Set 2), 0.9846 (Set 3) and 0.8317 (Set 4) & ANN with PCA &  The study separates basic activity (Set 1 and 2) data from transitional activity (Set 3 and 4), and then evaluate each set independently.\\ 
            
            Noor et al. \cite{MohdNoor2021} (2017) & SBHARPT (Acc): 0.934 & Decision tree with multivariate Gaussian & Distinguishing between non-transitional and transitional activities requires different models and an assumption about the data distribution.\\ 

            Baraka et al. \cite{Baraka2023} (2023) & SBHARPT (Acc): 0.927, FORTH-TRACE (Acc): 0.8665  & CNN & A non-end-to-end architecture is used, consisting of two separate classifiers: one for basic activities and the other for transitional activities.\\ 

            Irfan et al. \cite{Irfan2021} (2021) & SBHARPT (Acc): 0.961, Helou (Acc): 0.9838 & LSTM, BiLSTM and CNN & Ensemble learning is used to combine predictions from three different deep learning models\\ 

            Noor et al. \cite{MohdNoor2022} (2022) & SBHARPT (Acc): 0.916 & Hybrid CNN-LSTM & The model employs multiple feature learning pipelines to extract temporal features across adjacent windows.\\ 

            Baraka et al. \cite{Baraka2024} (2024) & SBHARPT (Acc): 0.934, FORTH-TRACE (Acc): 0.8496 & CNN & The method employs three input windows along with three dedicated pre-trained classifiers, respectively handling segmentation, basic activity recognition, and transitional activity recognition.\\ 

            Ige  et al. \cite{Ige2025} (2023) & PAMAP (Acc): 0.9852, WISDM (Acc): 0.9790 & Hybrid of CNN-BiLSTM with attention mechanism & The model employs multiple feature learning pipelines to capture local and temporal features and fuses them into a unified representation that captures complementary spatio-temporal representations. \\ 

            \textbf{This study} & SBHARPT: 0.9337 (Acc) / 0.8654 (F1), FORTH-TRACE: 0.8702 (Acc) / 0.7763 (F1) & JEPA-based convolutional and transformer models \\ 
            \hline
        \end{tabularx}
    \label{tab:compare_fully_supervised}
\end{sidewaystable}

As for the comparison with existing self-supervised learning approaches shown in Table \ref{tab:compare_self_supervised}, the majority of existing works rely on contrastive learning objectives, while only one study employs predictive pretext tasks based on masking strategies \cite{liu2025robusthar}. While these approaches achieve high performance, our approach utilizes a predictive objective in the latent space, offering a distinct paradigm for modeling human activity. Notably, among the listed methods, only Khaertdinov et al. \cite{khaertdinov_contrastive_2021} explicitly conduct a transfer learning setup, where pre-training is performed on MobiAct that differs from the downstream evaluation datasets. In contrast, the other approaches pre-train and evaluate the downstream activity recognition task using the same datasets, which may limit the assessment of cross-dataset generalization.

The contrastive approaches represented by Khaertdinov et al. \cite{khaertdinov_contrastive_2021} depends on time-domain transformation such as jittering, scaling, or rotation to create different perspective of the data, where dual augmented views are generated through random transformations and encoded using a hybrid CNN-Transformer architecture to capture both spatial and long-range temporal dependencies. Furthermore, Huang et al. \cite{huang_tfc_2025} proposed a temporal-frequency contrastive learning approach that generates both time and frequency-domain augmented views of sensor data, enabling the model to learn invariant and discriminative representations. While effective, these methods can inadvertently introduce noise or destroy intrinsic temporal and spectral patterns unique to specific activities. Furthermore, their dependence on negative sampling increases computational complexity and may lead to suboptimal representations when negative pairs are not truly dissimilar.

Recent contrastive learning approaches such as Chen et al. \cite{chen_temporal_2025} and Yarici et al. \cite{yarici_subject_2025} incorporate cross-modality fusion or subject-specific re-weighting to learn more robust and generalizable representations by either leveraging correlation across multiple sensor modalities or reducing the influence of subject-dependent variations. Although our accuracy is slightly lower, it should be noted that our evaluation was conducted in a transfer learning setup on datasets where sensor data was collected from continuously performed activities.

It is important to note that JEPA-based architectures are designed to ignore low-level sensor noise and focus on high-level modeling of the signal. Unlike generative self-supervised models that attempt to reconstruct raw signal values which are often stochastic in HAR. Also, unlike contrastive models also require large batches to distinguish between similar activities, JEPA framework learns stable, invariant features by operating entirely in the latent space. This suggests that the proposed method provides a more robust and scalable foundation for HAR, particularly in scenarios where data distribution is highly variable across subjects or sensor qualities.

\begin{sidewaystable}[p]
    \caption{Comparison between the proposed and self-supervised learning methods}    
        \centering
        \footnotesize
    \renewcommand{\arraystretch}{1.2}
    \setlength{\tabcolsep}{4pt}
    
    \begin{tabularx}{\textwidth}{>{\hsize=0.4\hsize}X >{\hsize=0.6\hsize}X >{\hsize=1.0\hsize}X >{\hsize=1.2\hsize}X >{\hsize=1.5\hsize}X}
            \hline
            Paper (Year) & Pre-training Dataset & Downstream Dataset (Metric) & Method & Description\\   
            \hline
            Liu et al. \cite{liu2025robusthar} (2025) & Opportunity, RealWorld, CZU-MHAD & Opportunity: 0.9372 (Acc) / 0.9374 (F1), RealWorld: 0.9576 (Acc) / 0.9578 (F1), CZU-MHAD: 0.9463 (Acc) / 0.9469 (F1) & \textbf{Encoder-Decoder}: Transformer & This generative self-supervised approach uses multi-scale spatial-temporal pre-training with segment, channel, and sensor masking. This allows a Transformer encoder-decoder to reconstruct masked content while maintaining semantic consistency across scales. \\
            
            Khaertdinov et al. \cite{khaertdinov_contrastive_2021} (2021) & MobiAct & UCI-HAR (F1): 0.8826, USC-HAD (F1): 0.4873 & \textbf{Augmentation}: jittering, scaling, channel shuffle, rotation and permutation. \textbf{Encoder}: CNN-Transformer & This contrastive learning strategy generates dual augmented views via random transformations and encodes them with a CNN-Transformer to capture spatial and long-range temporal features. It optimizes representations by maximizing cosine similarity for identical instances and minimizing it for disparate ones. \\

            Huang et al. \cite{huang_tfc_2025} (2025) & MotionSense, UCI-HAR, USC-HAD, BARD & MotionSense (Acc): 0.8984, UCI-HAR (Acc): 0.9535, USC-HAD (Acc): 0.8297, BARD (Acc): 0.9708 & \textbf{Augmentation}: rotation, translation and frequency perturbation. \textbf{Encoder}: temporal convolutional network & The method generates time- and frequency-domain augmented views of sensor signals, encodes them with a Temporal Convolutional Network, and applies contrastive learning both across predicted future time steps and frequency components to produce invariant, high-quality representations from unlabeled data. \\

            Chen et al. \cite{chen_temporal_2025} (2025) & MobiAct, UCI-HAR, USC-HAD & MobiAct (F1): 0.8688, UCI-HAR (F1): 0.9530, USC-HAD (F1): 0.6478 & \textbf{Encoder}: CNN-BiLSTM with attention mechanism & This contrastive learning method applies structured temporal and sensor-specific augmentations and encodes sequences with a CNN-BiLSTM-attention network. By projecting fused multi-sensor features into a latent space, it pulls embeddings of the same activity together and pushes different activities apart. \\

            Yarici et el. \cite{yarici_subject_2025} (2025) & UTD-MHAD, MMAct, DARai & UTD-MHAD (Acc): 0.9453, MMAct (Acc): 0.7497, DARai (Acc): 0.1615 & \textbf{Encoder}: CNN-Transformer & The method modifies standard contrastive learning by re-weighting negatives from the same subject, forcing embeddings to ignore subject-specific traits and focus on activity differences, producing representations that generalize better across unseen subjects. \\
            
            \textbf{This study} & REALDISP & SBHARPT: 0.9337 (Acc) / 0.8654 (F1), FORTH-TRACE: 0.8702 (Acc) / 0.7763 (F1) & \textbf{Augmentation}: channel reversal and noise addition. \textbf{Encoder}: Convolutional, Transformer \\ 
            \hline
        \end{tabularx}
    \label{tab:compare_self_supervised}
\end{sidewaystable}

\section{Conclusion}
In this paper, we present a JEPA framework for sensor-based HAR (HAR-JEPA) to learn robust and generalizable representations from unlabeled datasets. In the framework, the encoder is designed to explicitly model both temporal sequence of adjacent windows and local temporal representation within the windows. Furthermore, an improved VICReg objective is proposed by introducing a computationally lightweight norm term to stabilize the pre-training phase, and effectively preventing representation collapse. The proposed framework is evaluated using two benchmark HAR datasets. The results show that the proposed HAR-JEPA framework successfully learns high-quality representations that outperform or match fully supervised baselines and existing self-supervised approaches on continuously performed activities. In future work, the impact of data preprocessing such as window size, number of windows and sampling rate on the classification performance can be further explored.

\section*{Acknowledgements}
This work has been supported in part by the Ministry of Higher Education Malaysia for
Fundamental Research Grant Scheme with Project Code: FRGS/1/2023/ICT02/USM/02/2.

\section*{Declaration Statements}

\textbf{Ethics Approval}: Not applicable.

\textbf{Competing Interests}: The authors declare that they have no known competing financial interests or
personal relationships that could have appeared to influence the work reported in this paper.

\textbf{Data Availability}: The datasets used in this study are publicly available at UCI Machine Learning Repository.

\textbf{Code Availability}: Source code is available at \url{https://github.com/mohalim/JEPA_HAR/}

\printbibliography

@article{yuan_self_supervised_2024,
	title = {Self-supervised learning for human activity recognition using 700,000 person-days of wearable data},
	volume = {7},
	copyright = {2024 The Author(s)},
	issn = {2398-6352},
	url = {https://www.nature.com/articles/s41746-024-01062-3},
	doi = {10.1038/s41746-024-01062-3},
	language = {en},
	number = {1},
	urldate = {2026-01-22},
	journal = {npj Digital Medicine},
	author = {Yuan, Hang and Chan, Shing and Creagh, Andrew P. and Tong, Catherine and Acquah, Aidan and Clifton, David A. and Doherty, Aiden},
	month = apr,
	year = {2024},
	note = {Publisher: Nature Publishing Group},
	pages = {91},
}

@article{logacjov_selfpab_2024,
	title = {{SelfPAB}: large-scale pre-training on accelerometer data for human activity recognition},
	volume = {54},
	issn = {1573-7497},
	shorttitle = {{SelfPAB}},
	url = {https://doi.org/10.1007/s10489-024-05322-3},
	doi = {10.1007/s10489-024-05322-3},
	language = {en},
	number = {6},
	urldate = {2026-01-23},
	journal = {Applied Intelligence},
	author = {Logacjov, Aleksej and Herland, Sverre and Ustad, Astrid and Bach, Kerstin},
	month = mar,
	year = {2024},
	pages = {4545--4563}
}

@article{chen_temporal_2025,
	title = {Temporal {Contrastive} {Learning} for {Sensor}-{Based} {Human} {Activity} {Recognition}: {A} {Self}-{Supervised} {Approach}},
	volume = {25},
	issn = {1558-1748},
	shorttitle = {Temporal {Contrastive} {Learning} for {Sensor}-{Based} {Human} {Activity} {Recognition}},
	url = {https://ieeexplore.ieee.org/document/10755044},
	doi = {10.1109/JSEN.2024.3491933},
	number = {1},
	urldate = {2026-01-22},
	journal = {IEEE Sensors Journal},
	author = {Chen, Xiaobing and Zhou, Xiangwei and Sun, Mingxuan and Wang, Hao},
	month = jan,
	year = {2025},
	pages = {1839--1850}
}

@article{huang_tfc_2025,
	title = {{TFC}: {Time}–frequency contrasting network for wearable-based human activity recognition},
	volume = {319},
	issn = {0950-7051},
	shorttitle = {{TFC}},
	url = {https://www.sciencedirect.com/science/article/pii/S0950705125004204},
	doi = {10.1016/j.knosys.2025.113373},
	urldate = {2026-01-23},
	journal = {Knowledge-Based Systems},
	author = {Huang, Zhenzhen and Deng, Jiukai and Wang, Shengzhi and Tang, Chaogang and Xiao, Shuo},
	month = jun,
	year = {2025},
	pages = {113373}
}

@inproceedings{liu2025robusthar,
  title={RobustHAR: Multi-scale Spatial-temporal Masked Self-supervised Pre-training for Robust Human Activity Recognition},
  author={Liu, Xiao and Yuan, Guan and Zhang, Yanmei and Liu, Shang and Yan, Qiuyan},
  booktitle={Proceedings of the Thirty-Fourth International Joint Conference on Artificial Intelligence},
  pages={8563--8571},
  year={2025}
}

@article{Straczkiewicz2021,
	title = {A systematic review of smartphone-based human activity recognition methods for health research},
   	volume = {4},
	issn = {23986352},
	issue = {1},
   	doi = {10.1038/s41746-021-00514-4},
   	author = {Marcin Straczkiewicz and Peter James and Jukka Pekka Onnela},
   	journal = {npj Digital Medicine},
   	month = {12},
   	publisher = {Nature Research},
   	year = {2021}
}

@misc{assran_self_supervised_2023,
	title = {Self-{Supervised} {Learning} from {Images} with a {Joint}-{Embedding} {Predictive} {Architecture}},
	url = {http://arxiv.org/abs/2301.08243},
	doi = {10.48550/arXiv.2301.08243},
	urldate = {2026-01-24},
	publisher = {arXiv},
	author = {Assran, Mahmoud and Duval, Quentin and Misra, Ishan and Bojanowski, Piotr and Vincent, Pascal and Rabbat, Michael and LeCun, Yann and Ballas, Nicolas},
	month = apr,
	year = {2023},
	note = {arXiv:2301.08243 [cs]}
}

@misc{bardes_revisiting_2024,
	title = {Revisiting {Feature} {Prediction} for {Learning} {Visual} {Representations} from {Video}},
	url = {http://arxiv.org/abs/2404.08471},
	doi = {10.48550/arXiv.2404.08471},
	urldate = {2026-01-24},
	publisher = {arXiv},
	author = {Bardes, Adrien and Garrido, Quentin and Ponce, Jean and Chen, Xinlei and Rabbat, Michael and LeCun, Yann and Assran, Mahmoud and Ballas, Nicolas},
	month = feb,
	year = {2024},
	note = {arXiv:2404.08471 [cs]}
}

@misc{fei_jepa_2024,
	title = {A-{JEPA}: {Joint}-{Embedding} {Predictive} {Architecture} {Can} {Listen}},
	shorttitle = {A-{JEPA}},
	url = {http://arxiv.org/abs/2311.15830},
	doi = {10.48550/arXiv.2311.15830},
	urldate = {2026-01-22},
	publisher = {arXiv},
	author = {Fei, Zhengcong and Fan, Mingyuan and Huang, Junshi},
	month = jan,
	year = {2024},
	note = {arXiv:2311.15830 [cs]}
}

@misc{tuncay_audio-jepa_2025,
	title = {Audio-{JEPA}: {Joint}-{Embedding} {Predictive} {Architecture} for {Audio} {Representation} {Learning}},
	shorttitle = {Audio-{JEPA}},
	url = {http://arxiv.org/abs/2507.02915},
	doi = {10.48550/arXiv.2507.02915},
	urldate = {2026-01-22},
	publisher = {arXiv},
	author = {Tuncay, Ludovic and Labbé, Etienne and Benetos, Emmanouil and Pellegrini, Thomas},
	month = jun,
	year = {2025},
	note = {arXiv:2507.02915 [cs]}
}

@misc{weimann_self-supervised_2024,
	title = {Self-{Supervised} {Pre}-{Training} with {Joint}-{Embedding} {Predictive} {Architecture} {Boosts} {ECG} {Classification} {Performance}},
	url = {http://arxiv.org/abs/2410.13867},
	doi = {10.48550/arXiv.2410.13867},
	urldate = {2026-01-22},
	publisher = {arXiv},
	author = {Weimann, Kuba and Conrad, Tim O. F.},
	month = oct,
	year = {2024},
	note = {arXiv:2410.13867 [eess]}
}

@misc{kim_learning_2024,
	title = {Learning {General} {Representation} of 12-{Lead} {Electrocardiogram} with a {Joint}-{Embedding} {Predictive} {Architecture}},
	url = {http://arxiv.org/abs/2410.08559},
	doi = {10.48550/arXiv.2410.08559},
	urldate = {2026-01-22},
	publisher = {arXiv},
	author = {Kim, Sehun},
	month = dec,
	year = {2024},
	note = {arXiv:2410.08559 [cs]}
}

@misc{bardes_vicreg_2022,
	title = {{VICReg}: {Variance}-{Invariance}-{Covariance} {Regularization} for {Self}-{Supervised} {Learning}},
	shorttitle = {{VICReg}},
	url = {http://arxiv.org/abs/2105.04906},
	doi = {10.48550/arXiv.2105.04906},
	urldate = {2026-03-11},
	publisher = {arXiv},
	author = {Bardes, Adrien and Ponce, Jean and LeCun, Yann},
	month = jan,
	year = {2022},
	note = {arXiv:2105.04906 [cs]}
}

@misc{huang_bijepa_2026,
	title = {{BiJEPA}: {Bi}-directional {Joint} {Embedding} {Predictive} {Architecture} for {Symmetric} {Representation} {Learning}},
	shorttitle = {{BiJEPA}},
	url = {http://arxiv.org/abs/2603.00049},
	doi = {10.48550/arXiv.2603.00049},
	urldate = {2026-07-11},
	publisher = {arXiv},
	author = {Huang, Yongchao},
	month = feb,
	year = {2026},
	note = {arXiv:2603.00049 [cs.LG]},
}

@misc{kuang_radial-vcreg_2026,
	title = {Radial-{VCReg}: {More} {Informative} {Representation} {Learning} {Through} {Radial} {Gaussianization}},
	shorttitle = {Radial-{VCReg}},
	url = {https://arxiv.org/abs/2602.14272v1},
	language = {en},
	urldate = {2026-07-11},
	journal = {arXiv.org},
	author = {Kuang, Yilun and Dagade, Yash and Chakraborty, Deep and Learned-Miller, Erik and Balestriero, Randall and Rudner, Tim G. J. and LeCun, Yann},
	month = feb,
	year = {2026},
}

@misc{sepanj_kernel_2025,
	title = {Kernel {VICReg} for {Self}-{Supervised} {Learning} in {Reproducing} {Kernel} {Hilbert} {Space}},
	url = {https://arxiv.org/abs/2509.07289v2},
	doi = {10.3390/bdcc10030078},
	language = {en},
	urldate = {2026-07-11},
	journal = {arXiv.org},
	author = {Sepanj, M. Hadi and Ghojogh, Benyamin and Moradi, Saed and Fieguth, Paul},
	month = sep,
	year = {2025},
	doi = {10.3390/bdcc10030078},
}

@misc{loshchilov_decoupled_2019,
	title = {Decoupled {Weight} {Decay} {Regularization}},
	url = {http://arxiv.org/abs/1711.05101},
	doi = {10.48550/arXiv.1711.05101},
	urldate = {2026-03-16},
	publisher = {arXiv},
	author = {Loshchilov, Ilya and Hutter, Frank},
	month = jan,
	year = {2019},
	note = {arXiv:1711.05101 [cs]}
}

@dataset{realdisp,
	author = {Banos, Oresti and Tóth, Máté and Amft, Oliver},
	title= {REALDISP Activity Recognition Dataset},
  	year= {2012},
  	publisher = {UCI Machine Learning Repository},
  	doi= {https://doi.org/10.24432/C5GP6D}
}

@dataset{4th,
  author       = {Katerina Karagiannaki and
                  Athanasia Panousopoulou and
                  Panagiotis Tsakalides},
  title        = {The FORTH-TRACE dataset for human activity
                   recognition of simple activities and postural
                   transitions using a Body Area Network
                  },
  month        = jul,
  year         = 2016,
  publisher    = {Zenodo},
  doi          = {10.5281/zenodo.841301},
  url          = {https://doi.org/10.5281/zenodo.841301},
}

@dataset{sbharpt,
	title = {Transition-Aware Human Activity Recognition Using Smartphones},
	journal = {Neurocomputing},
	volume = {171},
	pages = {754-767},
	year = {2016},
	issn = {0925-2312},
	doi = {https://doi.org/10.1016/j.neucom.2015.07.085},
	url = {https://www.sciencedirect.com/science/article/pii/S0925231215010930},
	author = {Jorge-L. Reyes-Ortiz and Luca Oneto and Albert Samà and Xavier Parra and Davide Anguita},
    doi= {https://doi.org/10.24432/C54G7M}
}

@article{baraka_similarity_2023,
	title = {Similarity {Segmentation} {Approach} for {Sensor}-{Based} {Activity} {Recognition}},
	volume = {23},
	issn = {1558-1748},
	url = {https://ieeexplore.ieee.org/document/10188611},
	doi = {10.1109/JSEN.2023.3295778},
	number = {17},
	urldate = {2026-03-17},
	journal = {IEEE Sensors Journal},
	author = {Baraka, AbdulRahman M. A. and Mohd Noor, Mohd Halim},
	month = sep,
	year = {2023},
	pages = {19704--19716},
}

@article{baraka_deep_2025,
	title = {Deep similarity segmentation model for sensor-based activity recognition},
	volume = {84},
	issn = {1573-7721},
	url = {https://doi.org/10.1007/s11042-024-18933-2},
	doi = {10.1007/s11042-024-18933-2},
	language = {en},
	number = {11},
	urldate = {2026-03-17},
	journal = {Multimedia Tools and Applications},
	author = {Baraka, AbdulRahman and Mohd Noor, Mohd Halim},
	month = mar,
	year = {2025},
	pages = {8869--8892},
}

@article{Lone2021,
   author = {Kashif Javed Lone and Lal Hussain and Sharjil Saeed and Adil Aslam and Asim Maqbool and Faisal Mehmood Butt},
   doi = {10.1080/17455030.2021.1971325},
   issn = {17455049},
   journal = {Waves in Random and Complex Media},
   publisher = {Taylor and Francis Ltd.},
   title = {Detecting basic human activities and postural transition using robust machine learning techniques by applying dimensionality reduction methods},
   year = {2021}
}

@article{MohdNoor2021,
   author = {Mohd Halim Mohd Noor},
   doi = {10.1007/s00521-020-05638-4},
   issn = {14333058},
   issue = {17},
   journal = {Neural Computing and Applications},
   month = {9},
   pages = {10909-10922},
   publisher = {Springer Science and Business Media Deutschland GmbH},
   title = {Feature learning using convolutional denoising autoencoder for activity recognition},
   volume = {33},
   year = {2021}
}

@article{Baraka2023,
   author = {Abdulrahman M.A. Baraka and Mohd Halim Mohd Noor},
   doi = {10.1109/JSEN.2023.3295778},
   issn = {15581748},
   issue = {17},
   journal = {IEEE Sensors Journal},
   month = {9},
   pages = {19704-19716},
   publisher = {Institute of Electrical and Electronics Engineers Inc.},
   title = {Similarity Segmentation Approach for Sensor-Based Activity Recognition},
   volume = {23},
   year = {2023}
}

@article{Irfan2021,
   author = {Saad Irfan and Nadeem Anjum and Nayyer Masood and Ahmad S. Khattak and Naeem Ramzan},
   doi = {10.3390/s21248227},
   issn = {14248220},
   issue = {24},
   journal = {Sensors},
   month = {12},
   pmid = {34960321},
   publisher = {MDPI},
   title = {A novel hybrid deep learning model for human activity recognition based on transitional activities},
   volume = {21},
   year = {2021}
}

@article{MohdNoor2022,
   author = {Mohd Halim Mohd Noor and Sen Yan Tan and Mohd Nadhir Ab Wahab},
   doi = {10.1007/s11063-022-10799-5},
   issn = {1573773X},
   issue = {5},
   journal = {Neural Processing Letters},
   month = {10},
   pages = {4027-4049},
   publisher = {Springer},
   title = {Deep Temporal Conv-LSTM for Activity Recognition},
   volume = {54},
   year = {2022}
}

@article{Baraka2024,
   author = {Abdul Rahman Baraka and Mohd Halim Mohd Noor},
   doi = {10.1007/s11042-024-18933-2},
   issn = {15737721},
   journal = {Multimedia Tools and Applications},
   month = {3},
   publisher = {Springer},
   title = {Deep similarity segmentation model for sensor-based activity recognition},
   year = {2024}
}

@article{Ige2025,
   author = {Ayokunle Olalekan Ige and Daniel Ayo Oladele and Malusi Sibiya},
   doi = {10.3390/app152312661},
   issn = {20763417},
   issue = {23},
   journal = {Applied Sciences (Switzerland)},
   month = {12},
   publisher = {Multidisciplinary Digital Publishing Institute (MDPI)},
   title = {Retentive-HAR: Human Activity Recognition from Wearable Sensors with Enhanced Temporal and Inter-Feature Dependency Retention},
   volume = {15},
   year = {2025}
}

@article{Maddala2026,
   author = {Jeevan Babu Maddala and Shaheda Akthar},
   doi = {10.1007/s13198-026-03167-2},
   issn = {0976-4348},
   journal = {International Journal of System Assurance Engineering and Management},
   title = {Enhanced elderly activity recognition in smart home environments using ConvLSTM2D with localization},
   url = {https://doi.org/10.1007/s13198-026-03167-2},
   year = {2026}
}

@article{Etumusei2025,
   author = {Jonathan Etumusei and Jorge Martinez Carracedo and Sally McClean},
   doi = {10.1007/s42979-025-03895-5},
   issn = {26618907},
   issue = {5},
   journal = {SN Computer Science},
   month = {6},
   publisher = {Springer},
   title = {Enhancing Dynamic Human Activity Recognition Through a Novel Martingale-Based Algorithm for Change Detection},
   volume = {6},
   year = {2025}
}

@article{Zhang2026,
   author = {Song-Zhen Zhang and Hui-Zhen Yang and Yun Gao},
   doi = {10.1038/s41598-026-41195-x},
   issn = {2045-2322},
   journal = {Scientific Reports},
   title = {IoT framework for sports activity safety monitoring based on wearable sensors and CRNN spatiotemporal analysis},
   url = {https://doi.org/10.1038/s41598-026-41195-x},
   year = {2026}
}

@article{Haresamudram2025,
   author = {Harish Haresamudram and Chi Ian Tang and Sungho Suh and Paul Lukowicz and Thomas Plötz},
   doi = {10.1145/3729467},
   issn = {24749567},
   issue = {2},
   journal = {Proceedings of the ACM on Interactive, Mobile, Wearable and Ubiquitous Technologies},
   month = {6},
   publisher = {Association for Computing Machinery},
   title = {Past, Present, and Future of Sensor-based Human Activity Recognition Using Wearables: A Surveying Tutorial on a Still Challenging Task},
   volume = {9},
   year = {2025}
}

@inproceedings{khaertdinov_contrastive_2021,
	title = {Contrastive {Self}-supervised {Learning} for {Sensor}-based {Human} {Activity} {Recognition}},
	issn = {2474-9699},
	url = {https://ieeexplore.ieee.org/document/9484410},
	doi = {10.1109/IJCB52358.2021.9484410},
	urldate = {2026-01-22},
	booktitle = {2021 {IEEE} {International} {Joint} {Conference} on {Biometrics} ({IJCB})},
	author = {Khaertdinov, Bulat and Ghaleb, Esam and Asteriadis, Stylianos},
	month = aug,
	year = {2021},
	pages = {1--8}
}

@misc{yarici_subject_2025,
	title = {Subject {Invariant} {Contrastive} {Learning} for {Human} {Activity} {Recognition}},
	url = {http://arxiv.org/abs/2507.03250},
	doi = {10.48550/arXiv.2507.03250},
	urldate = {2026-01-22},
	publisher = {arXiv},
	author = {Yarici, Yavuz and Kokilepersaud, Kiran and Prabhushankar, Mohit and AlRegib, Ghassan},
	month = jul,
	year = {2025},
	note = {arXiv:2507.03250 [cs]}
}

@article{Shi2021,
		author = {Shi, Junhao and Zuo, Decheng and Zhang, Zhan},
		title = {A GAN-based data augmentation method for human activity recognition via the caching ability},
		journal = {Internet Technology Letters},
		volume = {4},
		number = {5},
		pages = {e257},
		doi = {https://doi.org/10.1002/itl2.257},
		url = {https://onlinelibrary.wiley.com/doi/abs/10.1002/itl2.257},
		eprint = {https://onlinelibrary.wiley.com/doi/pdf/10.1002/itl2.257},
		year = {2021}
}

@Article{Zhang2022,
		AUTHOR = {Zhang, Shibo and Li, Yaxuan and Zhang, Shen and Shahabi, Farzad and Xia, Stephen and Deng, Yu and Alshurafa, Nabil},
		TITLE = {Deep Learning in Human Activity Recognition with Wearable Sensors: A Review on Advances},
		JOURNAL = {Sensors},
		VOLUME = {22},
		YEAR = {2022},
		NUMBER = {4},
		ARTICLE-NUMBER = {1476},
		URL = {https://www.mdpi.com/1424-8220/22/4/1476},
		PubMedID = {35214377},
		ISSN = {1424-8220},
		DOI = {10.3390/s22041476}
}

\end{document}